%% file: TOP-17-012_temp.tex
\pdfoutput=1

\documentclass[11pt,twoside,a4paper,cmspaper,final,collab]{cms-tdr}
\def\svnVersion{6cf7b83-D}\def\svnDate{2020/07/22}\def\cmsCernNoTag{CERN-EP-2020-047}\def\cmsCernDate{\today}\def\cmsMessage{"Published in Physics Letters B as \href{http://dx.doi.org/10.1016/j.physletb.2020.135609}{\doi{10.1016/j.physletb.2020.135609}.}"}

\begin{document}\cmsNoteHeader{TOP-17-012}

\hyphenation{had-ron-i-za-tion}
\hyphenation{cal-or-i-me-ter}
\hyphenation{de-vices}
\RCS$HeadURL$
\RCS$Id$
\newlength\cmsFigWidth
\ifthenelse{\boolean{cms@external}}{\setlength\cmsFigWidth{0.85\columnwidth}}{\setlength\cmsFigWidth{0.4\textwidth}}
\ifthenelse{\boolean{cms@external}}{\providecommand{\cmsLeft}{top\xspace}}{\providecommand{\cmsLeft}{left\xspace}}
\ifthenelse{\boolean{cms@external}}{\providecommand{\cmsRight}{bottom\xspace}}{\providecommand{\cmsRight}{right\xspace}}

\newcommand{\etalj}{\ensuremath{\eta_{\text{j}'}}\xspace}
\newcommand{\abseta}{\ensuremath{\abs{\eta}}\xspace}
\newcommand{\mpx}{\ensuremath{p_{x}^{\text{miss}}}\xspace}
\newcommand{\mpy}{\ensuremath{p_{y}^{\text{miss}}}\xspace}
\newcommand{\mtw}{\ensuremath{\mT^{\PW}}\xspace}
\newcommand{\mW}{\ensuremath{m_{\PW}}\xspace}
\newcommand{\QCD}{\ensuremath{\mathrm{QCD}~\text{multijet}}\xspace}
\newcommand{\wjets}{\ensuremath{\PW\text{+jets}}\xspace}
\newcommand{\mt}{\ensuremath{m_{\ell\nu \cPqb}}\xspace}
\newcommand{\vtb}{\ensuremath{V_{\PQt\PQb}}\xspace}
\newcommand{\vtd}{\ensuremath{V_{\PQt\PQd}}\xspace}
\newcommand{\vts}{\ensuremath{V_{\PQt\PQs}}\xspace}
\newcommand{\absvtb}{\ensuremath{\abs{V_{\PQt\PQb}}}\xspace}
\newcommand{\absvts}{\ensuremath{\abs{V_{\PQt\PQs}}}\xspace}
\newcommand{\absvtd}{\ensuremath{\abs{V_{\PQt\PQd}}}\xspace}
\newcommand{\absvtq}{\ensuremath{\abs{V_{\PQt\PQq}}}\xspace}
\newcommand{\absvtbsq}{\ensuremath{\abs{V_{\PQt\PQb}}^2}\xspace}
\newcommand{\absvtssq}{\ensuremath{\abs{V_{\PQt\PQs}}^2}\xspace}
\newcommand{\absvtdsq}{\ensuremath{\abs{V_{\PQt\PQd}}^2}\xspace}
\newcommand{\absvtqsq}{\ensuremath{\abs{V_{\PQt\PQq}}^2}\xspace}
\newcommand{\absvtbfourth}{\ensuremath{\abs{V_{\PQt\PQb}}^4}\xspace}
\newcommand{\sigmatq}{\ensuremath{\sigma_{t\text{-ch,}\PQq}\xspace}}
\newcommand{\TTqb}{\ensuremath{\ttbar_{\PQb,\PQq}\xspace}}
\newcommand{\STbb}{\ensuremath{ST_{\PQb,\PQb}\xspace}}
\newcommand{\STqb}{\ensuremath{ST_{\PQq,\PQb}\xspace}}
\newcommand{\STbq}{\ensuremath{ST_{\PQb,\PQq}\xspace}}
\newcommand{\sigmatb}{\ensuremath{\sigma_{t\text{-ch,}\PQb}\xspace}}
\newcommand{\BRtb}{\ensuremath{\mathcal{B}(\PQt\to \PW\PQb)\xspace}}
\newcommand{\BRtq}{\ensuremath{\mathcal{B}(\PQt \to \PW\PQq)\xspace}}
\newcommand{\sigmats}{\ensuremath{\sigma_{t\text{-ch},\PQs}\xspace}}
\newcommand{\BRts}{\ensuremath{\mathcal{B}(\PQt \to \PW\PQs)\xspace}}
\newcommand{\sigmatd}{\ensuremath{\sigma_{t\text{-ch,}\PQd}\xspace}}
\newcommand{\sigmatsd}{\ensuremath{\sigma_{t\text{-ch},\PQs,\PQd}\xspace}}
\newcommand{\BRtsd}{\ensuremath{\mathcal{B}(\PQt \to \PW\PQs,\PQd)\xspace}}
\newcommand{\twoJoneT}{2j1t\xspace}
\newcommand{\threeJoneT}{3j1t\xspace}
\newcommand{\threeJtwoT}{3j2t\xspace}

\newlength\cmsTabSkip\setlength{\cmsTabSkip}{1ex}
\providecommand{\CL}{CL\xspace}

\cmsNoteHeader{TOP-17-012}
\title{Measurement of CKM matrix elements in single top quark $t$-channel production in proton-proton collisions at $\sqrt{s} = 13\TeV$}
\date{\today}

\abstract{
The first direct, model-independent measurement is presented of the modulus of the Cabibbo--Kobayashi--Maskawa (CKM) matrix elements \absvtb, \absvtd, and \absvts, in final states enriched in single top quark $t$-channel events. 
The analysis uses proton-proton collision data from the LHC, collected during 2016 by the CMS experiment, at a centre-of-mass energy of 13\TeV, corresponding to an integrated luminosity of 35.9\fbinv.
Processes directly sensitive to these matrix elements are considered at both the production and decay vertices of the top quark.
In the standard model hypothesis of CKM unitarity, a lower limit of $\absvtb > 0.970$ is measured at the 95$\%$ confidence level. Several theories beyond the standard model are considered, and by releasing all constraints among the involved parameters, the values $\absvtb = 0.988 \pm 0.024 $, and $\absvtdsq +\absvtssq = 0.06 \pm 0.06$, where the uncertainties include both statistical and systematic components, are measured.}
\hypersetup{
pdfauthor={CMS Collaboration},
pdftitle={Measurement of CKM matrix elements in single top quark t-channel events at sqrt(s) = 13 TeV},
pdfsubject={CMS},
pdfkeywords={CMS, physics, CKM matrix, top quark}}
\maketitle
\section{Introduction}
\label{sec:introduction}
\begin{figure*}[h]
\centering
\includegraphics[width=0.46\textwidth]{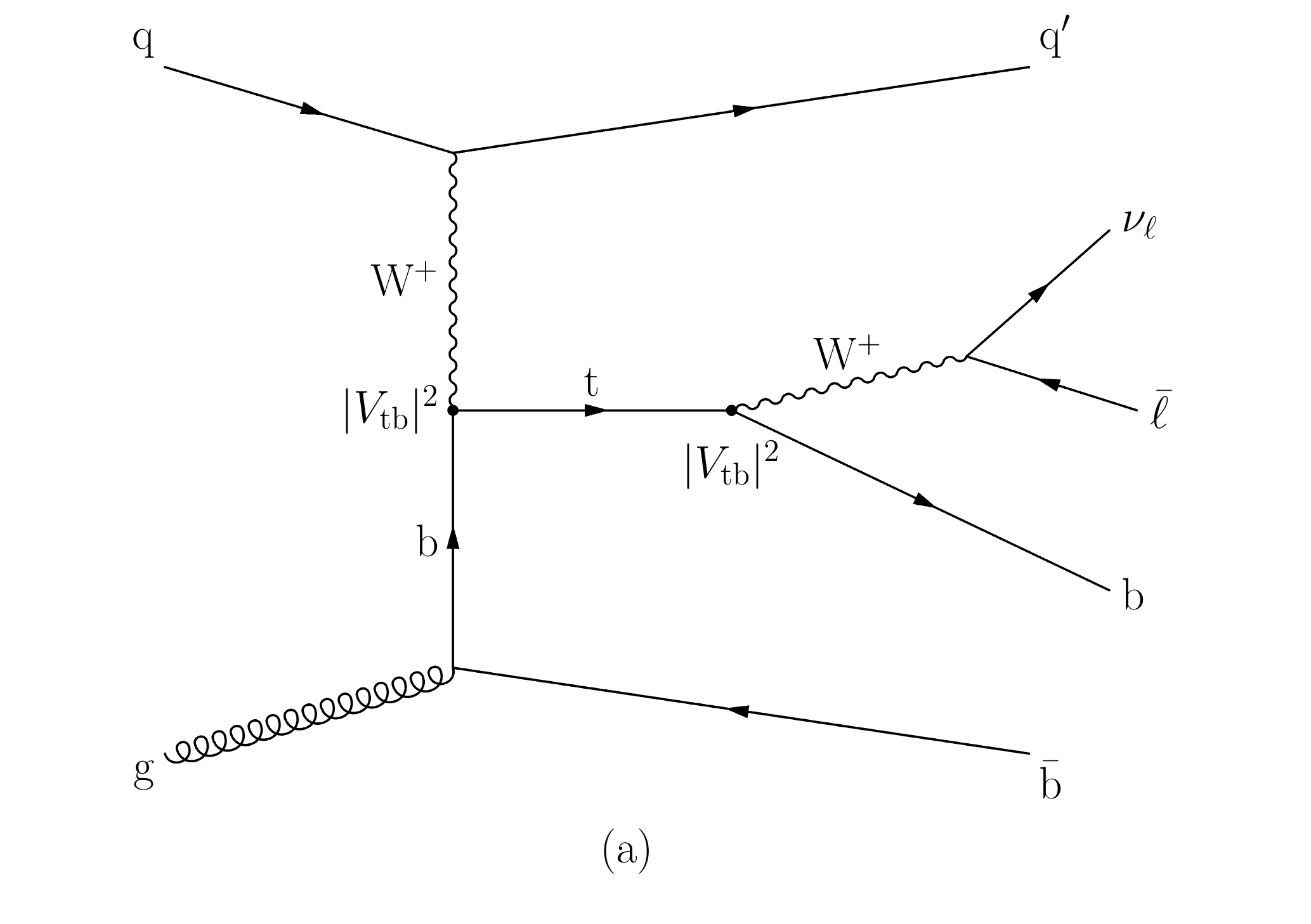}\,\,
\includegraphics[width=0.46\textwidth]{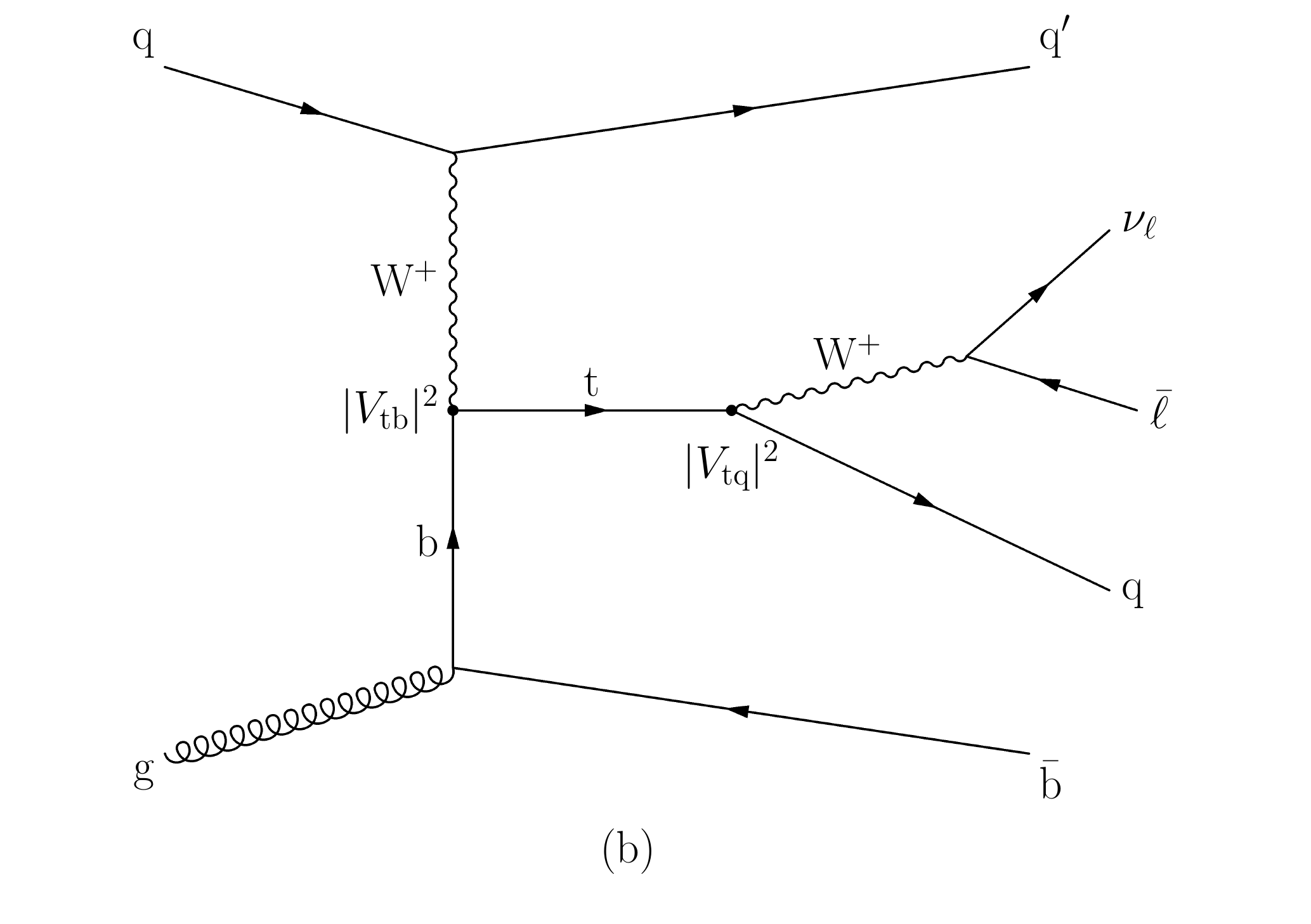}\,\,
\includegraphics[width=0.46\textwidth]{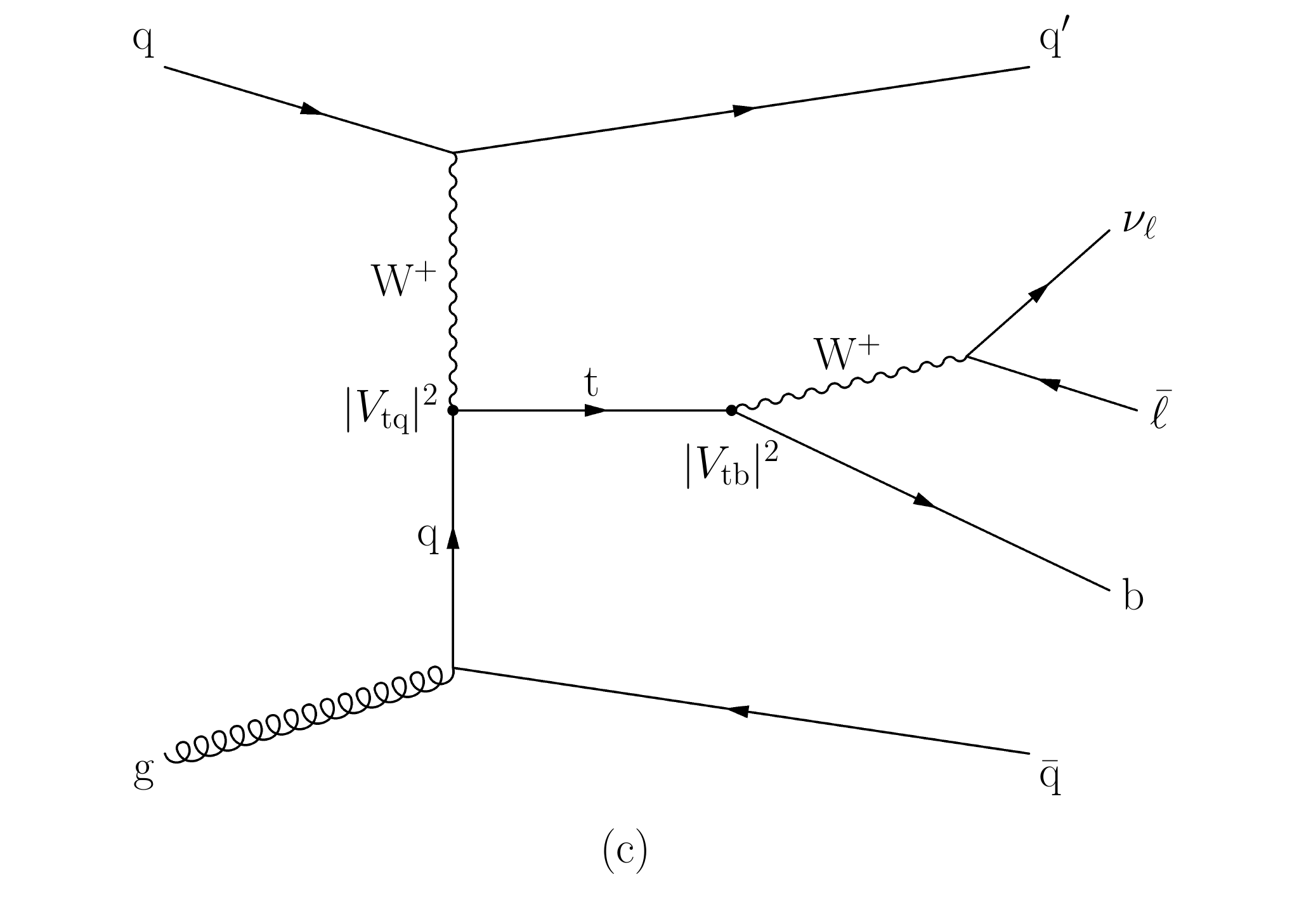}\,\,
\includegraphics[width=0.46\textwidth]{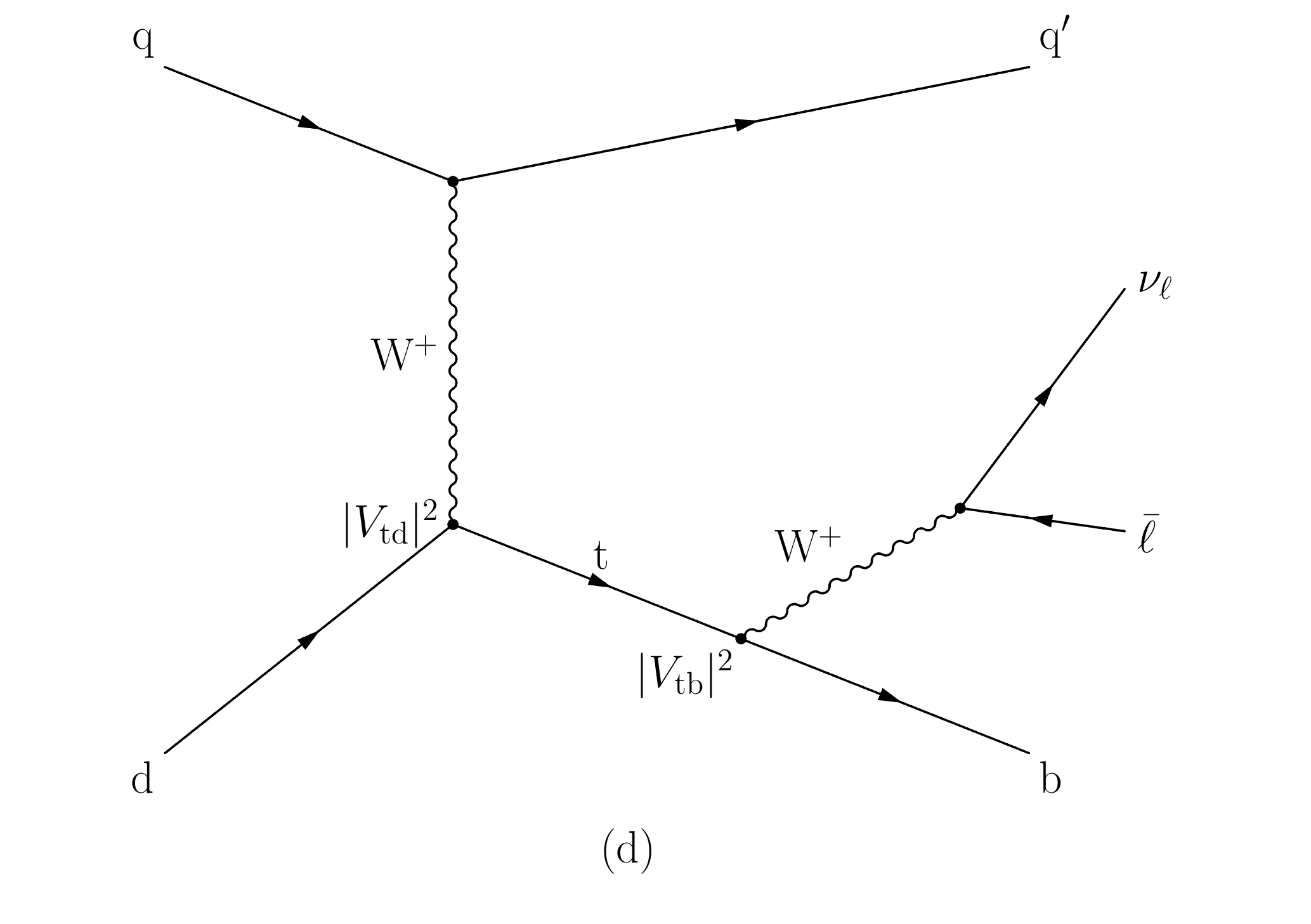}\,\,
\caption{\label{fig:FG} Leading-order Feynman diagrams for single top quark production via the $t$ channel featuring: (a) a $\PQt\PW\PQb$ vertex in production and decay, (b) a $\PQt\PW\PQb$ vertex in production and a $\PQt\PW\PQq$ in decay, with \PQq being an \PQs or \PQd quark, (c) a $\PQt\PW\PQq$ vertex in production and a $\PQt\PW\PQb$ in decay, and (d) a process initiated by a \PQd quark and enhanced due to contributions from these valence quarks. The $\ell$ refers to \Pe or \PGm leptons.}
\end{figure*}

A distinctive feature of the electroweak sector of top quark physics is the relative magnitude of the Cabibbo--Kobayashi--Maskawa (CKM)~\cite{Kobayashi:1973fv} matrix element $\vtb$ with respect to $\vtd$ and $\vts$, which leads to a strong suppression of processes involving mixing between the third and the first two quark families. 
This feature can be probed at the CERN LHC by studying the couplings of top quarks to \PQd, \PQs, and \PQb quarks in electroweak charged-current interactions, where such couplings play a role at either the production or decay vertices of the top quark. 
In general, top quarks are produced in proton-proton ($\Pp\Pp$) collisions through the strong interaction, predominantly via gluon fusion, creating a top quark-antiquark (\ttbar) pair. Top quarks can also be singly produced via the electroweak interaction, in which case the dominant mechanism involves an exchange of a \PW boson in the $t$ channel, a process which has been precisely measured at the LHC~\cite{Chatrchyan:2011vp,Chatrchyan:2012ep,Aad:2012ux,Aad:2014fwa,Khachatryan:2014iya,Aaboud:2016ymp,Sirunyan:2016cdg,Aaboud:2017pdi,Sirunyan:2018rlu,Sirunyan:2019hqb}.
The dominant decay process for a top quark is to a \PW boson and a \PQb quark via an electroweak charged-current interaction.
All single top quark processes therefore allow the direct probing of the $\PQt\PW\PQq$ vertex, with \PQq representing a \PQb, \PQd, or \PQs quark, both in production and decay of the top quark.
In the $t$ channel, a top quark is produced recoiling against a light quark, henceforth referred to as $\PQq'$.  Figure~\ref{fig:FG} shows typical Feynman diagrams at leading order (LO) for the different production and decay modes considered in this analysis.

The elements $\vtb$, $\vtd$, and $\vts$ of the CKM matrix can be indirectly constrained from measurements in the \PB and \PK meson sectors~\cite{PDG2018}, but those determinations rely crucially on model assumptions, such as the existence of only three generations of quarks and the absence of particles beyond the standard model (SM)~\cite{Alwall:2006bx}. 
This model dependence motivates alternative inferences based on different sets of hypotheses. 
In particular, given that these three CKM elements connect the top quark with down-type quarks, it is natural to use events enriched in top quarks to set constraints on them. 
Two complementary approaches have been pursued by the Tevatron and LHC experiments to extract $\absvtb$: the first method measures the branching fraction $\BRtb = \absvtbsq / \bigl(\absvtdsq + \absvtssq + \absvtbsq\bigr)$ in \ttbar events~\cite{Abazov:2011zk,Aaltonen:2013luz,Aaltonen:2014yua,Khachatryan:2014nda}.
The $\BRtb$ measurement is sensitive to the CKM elements of interest through the decay vertex of the top quark, and can be turned into a measurement of the value of $\absvtb$ only under the hypothesis of the unitarity of the $3\times 3$ CKM matrix. 
The second method is based on the single top quark production cross section and is sensitive in principle through both the production and decay of the top quark. 
To disentangle the effects of the two vertices in past measurements at the Tevatron~\cite{Abazov:2006gd,Abazov:2008kt,Aaltonen:2008sy,Aaltonen:2009jj,Aaltonen:2010jr,Abazov:2009ii,Group:2009qk,Aaltonen:2014ura,Abazov:2013qka,CDF:2014uma} and the LHC~\cite{Aad:2014fwa,Chatrchyan:2011vp,Chatrchyan:2012ep,Aaboud:2017pdi,Khachatryan:2014iya,Aaboud:2016ymp,Sirunyan:2016cdg,Khachatryan:2016sib,Aaboud:2019pkc,Sirunyan:2018rlu}, $\absvtb$ was extracted in the $t$ channel by assuming that the values of $\absvts$ and $\absvtd$ are negligible. 
Some theoretical proposals have suggested the simultaneous extraction of the three CKM matrix elements from a combination of measurements of $\BRtb$ and either inclusive~\cite{Alwall:2006bx,Lacker:2012ek} or differential~\cite{AguilarSaavedra:2010wf,Clerbaux:2018vup} cross sections of single top quark production in the $t$ channel.
Other studies specifically address the determination of $\absvtd$~\cite{AguilarSaavedra:2010wf,Alvarez:2017ybk} through a reliance on the reinterpretation of existing measurements, but they do not make use of full experimental detector simulations and do not exploit the discriminating power of multivariate analyses (see, for example, discussion in Ref.~\cite{Giammanco:2017xyn}).

The data used in this Letter come from $\Pp\Pp$ collisions at $\sqrt{s}=13\TeV$, corresponding to an integrated luminosity of $35.9\fbinv$ and collected by the CMS experiment with triggers requiring either one muon or electron in the final state.
We present the first direct and model-independent simultaneous measurement of $\absvtb$, $\absvtd$, and $\absvts$, by considering their respective contributions to the top quark $t$-channel production and decay.

\section{The CMS detector}
\label{sec:cms}

The central feature of the CMS apparatus is a superconducting solenoid of 6\unit{m} internal diameter, providing a magnetic field of 3.8\unit{T}. Within the solenoid volume are a silicon pixel and strip tracker, a lead tungstate crystal electromagnetic calorimeter, and a brass and scintillator hadron calorimeter, each composed of a barrel and two endcap sections. Forward calorimeters extend the pseudorapidity ($\eta$)~\cite{Chatrchyan:2008zzk} coverage provided by the barrel and endcap detectors. 

Events of interest are selected using a two-tiered trigger system~\cite{Khachatryan:2016bia}. The first level, composed of custom hardware processors, uses information from the calorimeters and muon detectors to select events. The second level, known as the high-level trigger, consists of a farm of processors running a version of the full event reconstruction software optimised for efficient processing.
A more detailed description of the CMS detector, together with a definition of the coordinate system used and the relevant kinematic variables, can be found in Ref.~\cite{Chatrchyan:2008zzk}. 

\section{Simulated samples}
\label{sec:samples}

Monte Carlo (MC) event generators are used to simulate signal and background samples. Single top quark $t$-channel events are generated at next-to-leading-order (NLO) in quantum chromodynamics (QCD) with \POWHEG~2.0~\cite{Nason:2004rx,Frixione:2007vw,Alioli:2010xd}. The four-flavour scheme~\cite{fourfiveflavorschemes} is used for events with the $\vtb$ vertex in production, while the five-flavour scheme~\cite{Alioli:2009je} is used for events with one $\vtd$ or $\vts$ vertex in production. Top quark decays are simulated with \textsc{madspin}~\cite{Artoisenet:2012st}. The $\ttbar$ background process~\cite{Frixione:2007nw},  as well as double vector boson production~\cite{Melia:2011tj,Nason:2013ydw} ({\PV\PV}, where \PV stands for either a \PW or a \PZ boson), are generated with \POWHEG~2.0. Associated top quark and \PW boson production are simulated with \POWHEG in the five-flavour scheme~\cite{Re:2010bp}. Single top quark $s$-channel events (t, $s$-ch) are simulated with \MGvATNLO~2.2.2~\cite{amcatnlo} at NLO.
The value of the top quark mass used in the simulated samples is 172.5\GeV. 
For all samples {\PYTHIA~8.180}~\cite{Sjostrand:2007gs} with tune CUETP8M1~\cite{Ctune} is used to simulate the parton shower, quark hadronisation, and underlying event, except for $\ttbar$, where the tune {CUETPM2T4} is used~\cite{CMS:2016kle}.
Simulated event samples with \PW and \PZ bosons in association with jets (\wjets, $\PZ$+jets) are generated using \MGvATNLO~2.2.2. For these processes, events with up to two additional partons emitted in the hard scattering are simulated, and the FxFx merging scheme~\cite{Frederix:2012ps} is used to avoid double counting with parton emissions generated in the parton showering. Simulated QCD multijet events, generated at LO with {\PYTHIA~8.180}, are used to validate the estimation of this background with a technique based on control samples in data.

The default parametrisation of the parton distribution functions (PDFs) used in all simulations is NNPDF3.0~\cite{NNPDF30} at LO or NLO QCD, with the order matching that of the matrix element calculation. All generated events undergo a full simulation of the detector response according to the model of the CMS detector within {\GEANTfour}~\cite{Agostinelli:2002hh}. Additional $\Pp\Pp$ interactions within the same or nearby bunch crossings (pileup) are included in the simulation with the same distribution as observed in data. Except for the QCD multijet process, which is determined from a fit to data, all simulated samples are normalised to the expected cross sections.

\section{Event selection and reconstruction}
\label{sec:selection}

The signal event selection is based on final states where the top quark decays to a \PQb, \PQs, or \PQd quark, and a \PW boson, which then decays to a lepton-neutrino pair. Events with exactly one muon or electron and at least two jets are considered in this analysis, as was done in the latest CMS single top quark cross section measurement~\cite{Sirunyan:2018rlu}. The neutrino accompanying the lepton cannot be directly detected, and manifests itself in the detector as a measured momentum imbalance in the event. Depending on the CKM matrix element involved in the decay, the final state may include a jet from the hadronisation of either a \PQb, \PQs, or \PQd quark. A jet recoiling against the top quark is present, and it is produced usually at low angle with respect to the beam axis. A third jet can stem from the second quark produced in the gluon splitting (as shown in Fig.~\ref{fig:FG}(c)). 
The quark from gluon splitting generates a jet that usually has a softer transverse momentum (\pt) spectrum than that of the jet from the top quark decay products. 
Depending on the number of $\PQt\PW\PQb$ vertices in the event process, one can have one jet coming from the hadronisation of a \PQb quark (\PQb jet) if the $\PQt\PW\PQb$ vertex occurs in production or in decay but not in both, or two \PQb jets if the $\PQt\PW\PQb$ vertex occurs both in production and decay.

Events are retained for the offline analysis if they were selected online by requiring the presence of an isolated, high-$\pt$ lepton: either a muon with $\pt > 24\GeV$ or an electron with $\pt>32\GeV$.
From the sample of triggered events, only those with at least one primary vertex reconstructed from at least four tracks, with a longitudinal distance of less than 24\unit{cm} and a radial distance of less then 2\unit{cm} from the centre of the detector, are considered for the analysis. 
The candidate vertex with the largest value of summed physics-object $\pt^2$ is taken to be the primary $\Pp\Pp$ interaction vertex. The physics objects are the jets, clustered using the jet-finding algorithm~\cite{fastjet2,Cacciari:2008gp} with the tracks assigned to candidate vertices as inputs, and the associated missing \pt (\ptmiss), taken as the magnitude of the negative vector sum of the \ptvecmiss of those jets.

 The particle-flow (PF) algorithm~\cite{PFPAPER} is used to reconstruct and identify individual particles in the event using combined information from the subdetectors of the CMS experiment, allowing identification of muons, electrons, photons, and charged and neutral hadrons. After triggering, muons are considered for further analysis if they have $\pt >26\GeV$ and $\abs{\eta} <2.4$, while electrons are required to have $\pt >35\GeV$ and $\abs{\eta}<2.1$.
Additional isolation requirements are used to discriminate between prompt leptons and those coming from hadronic decays within jets, by defining $I_{rel}$, as the scalar sum of the \pt of charged hadrons, neutral hadrons, and photons divided by the \pt of lepton in a cone of $\Delta R = \sqrt{\smash[b]{(\Delta\eta)^2+(\Delta\phi)^2}} = 0.4$ around the muon and 0.3 around the electron, where $\phi$ is the azimuthal angle in radians. The contribution of hadrons from pileup interactions is subtracted from the scalar sum with the techniques detailed in Refs.~\cite{Sirunyan:2018fpa,Cacciari:2007fd}. The parameter $I_{rel}$ is required to be less than 6.0\% for muons, 5.9\% for barrel electrons, and 5.7\% for endcap electrons.

Jets are reconstructed using the anti-\kt clustering algorithm described in Refs.~\cite{fastjet2,Cacciari:2008gp} with a distance parameter of 0.4 on the collection of PF candidates.
To be included, charged particle candidates must be closer along the z axis to the primary vertex than to any other vertex.

A correction to account for pileup interactions is estimated on an event-by-event basis using the jet area method described in Ref.~\cite{Cacciari:2007fd}, and is applied to the reconstructed jet \pt. Further jet energy corrections~\cite{Khachatryan:2016kdb}, derived from the study of dijet events and photon plus jet events in data, are applied. Two types of jets are defined: high-\pt jets are defined by requiring $\abs{\eta}<4.7$ and $\pt > 40~\GeV$, and  low-\pt jets are defined by requiring $\abs{\eta}<4.7$ and $20 < \pt < 40~\GeV$.

Once the jets have been selected according to the above criteria, they can be further categorised using a \PQb tagging discriminator variable in order to distinguish between jets stemming from the hadronisation of \PQb quarks and those from the hadronisation of light partons. A multivariate (MVA) discriminator algorithm uses track-based lifetime information, together with secondary vertices inside the jet, to provide a MVA discriminator for \PQb jet identification~\cite{1748-0221-8-04-P04013,BTAGPAPER}. For values of the discriminator above the chosen threshold, the efficiency of the tagging algorithm to correctly find \PQb jets is about 45\%, with a rate of 0.1\% for mistagging light-parton jets~\cite{1748-0221-8-04-P04013,BTAGPAPER}.

Events are divided into mutually exclusive ``categories'' according to the number of selected high-\pt jets and \PQb-tagged high-\pt jets. In the following, categories are labelled as ``$n$j$m$t'', referring to events with exactly $n$ high-\pt jets, $m$ of which are tagged as \PQb jets, regardless of the number of low-\pt jets. The threshold on the jet momentum for high-\pt jets lessens the impact on the categorisation of additional jets coming from initial- or final-state radiation, which is fully simulated and taken into account in the modelling systematic uncertainties.

To reject events from QCD multijet background processes, a requirement on the transverse mass of the \PW boson of $\mtw > 50\GeV$ is imposed, where
\begin{linenomath}
\begin{equation}
\mtw =\sqrt{\left(p_{\mathrm{T},\ell} + \ptmiss \right)^2 - \left( p_{x,\ell} + \mpx \right)^2 - \left( p_{y,\ell} + \mpy \right)^2}.
\end{equation}
\end{linenomath}
Here, $\ptmiss$ is defined as the magnitude of $\ptvecmiss$, which is the negative of the vectorial \ptvec sum of all the PF particles. The $\mpx$ and $\mpy$  quantities are the $\ptvecmiss$ components along the $x$ and $y$ axes, respectively, and $p_{\mathrm{T},\ell}$, $p_{x,\ell}$, and $p_{y,\ell}$ are the corresponding lepton momentum components in the transverse, $x$, and $y$ directions.

To analyse the kinematics of single top quark production, the four-momentum of a top quark candidate is reconstructed from the decay products: leptons, neutrinos, and \PQb jet candidates. The \pt of the neutrino can be inferred from \ptmiss. The longitudinal momentum of the neutrino, $p_{z,\PGn}$, is calculated assuming energy-momentum conservation at the $\PW\ell\PGn$ vertex and constraining the \PW boson mass to $m_{\mathrm{W}}  = 80.4\GeV$~\cite{PDG2018}:
\begin{linenomath}
\begin{equation}
\label{eq:nusolver}
p_{z,\PGn}^{\pm} =\frac{\Lambda p_{z,\ell}}{p_{\mathrm{T},\ell}^2}\pm\frac{1}{p_{\mathrm{T},\ell}^2}\sqrt{\Lambda^2 p_{z,\ell}^2-p_{\mathrm{T},\ell}^2(p_{\ell}^{2} p_{\mathrm{T},\PGn}^2-\Lambda^2)},
\end{equation} 
\end{linenomath}
where
\begin{linenomath}
\begin{equation}
\label{eq:lambda}
\Lambda=\frac{m_{\PW}^2}{2}+\vec{p}_{\mathrm{T},\ell}\cdot\ptvecmiss,
\end{equation} 
\end{linenomath}
and $p_{\ell}^{2}=p_{\mathrm{T},\ell}^2 + p_{z,\ell}^2$ denotes the square of the lepton momentum.
In most of the cases, this leads to two real solutions for $p_{z,\PGn}$ and  the solution with the smallest absolute value is chosen~\cite{Abazov:2009ii,Aaltonen:2009jj}. For some events, the discriminant in Eq.~(\ref{eq:nusolver}) becomes negative, leading to complex solutions for $p_{z,\PGn}$. In this case, the imaginary component is eliminated by modification of $p_{x,\PGn}$ and $p_{y,\PGn}$ so that $\mtw = \mW$, while still respecting the $\mW$ constraint. This is achieved by requiring the determinant, and thus the square-root term in Eq.~(\ref{eq:nusolver}), to equal zero. This condition gives a quadratic relation between $p_{x,\PGn}$ and $p_{y,\PGn}$ with two possible solutions and one remaining degree of freedom. The solution is chosen by finding the $\vec{p}_{\mathrm{T},\PGn}$  that has the minimum vectorial distance from \ptvecmiss in the $\mpx - \mpy$ plane. 

A reconstructed top quark candidate is defined by associating one jet with an accompanying \PW boson, and the respective top quark four-momentum is evaluated as described above. 
For each of the signal categories selected, multiple top quark candidates can be defined in the same category, depending on the hypothesis for the origin of the jet in the event.

\section{Signal description and event categorisation}
\label{sec:ckmdiscr}

The predicted branching fractions of top quarks to \PQd, \PQs, and \PQb quarks can be written as a function of the overall magnitude of $\BRtq = \absvtqsq / \bigl(\absvtdsq + \absvtssq + \absvtbsq\bigr)$. The values of $\absvtq$ and $\BRtq$ used to derive the initial normalisation of signal processes are taken from Ref.~\cite{PDG2018}, and shown in Table~\ref{tab:matrix}.

\begin{table*}[ht]
 \topcaption{Values of the third-row elements of the CKM matrix inferred from low-energy measurements, taken from Ref.~\cite{PDG2018}, with the respective values of the top quark decay branching fractions. The \PQq in \absvtq and $\BRtq$ in the first column refers to \PQb, \PQs, and \PQd quarks, according to the quark label shown in the header row.} 
\label{tab:matrix}  
\centering
{\renewcommand\arraystretch{1.3} 
\begin{tabular}{ lccc }
Quark & \PQb & \PQs & \PQd\\
\hline
\absvtq & $0.999119^{\,+0.000024}_{\,-0.000012}$ & $0.04108^{\,+0.00030}_{\,-0.00057}$ & $0.008575^{\,+0.000076}_{\,-0.000098}$  \\
\BRtq  & $0.998239^{\,+0.000048}_{\,-0.000024}$ & $0.0016876^{\,+0.0000025}_{\,-0.0000047}$ & $0.000074^{\,+0.000013}_{\,-0.000017}$\\
\end{tabular} 
}
\end{table*} 

The quantities reported in Table~\ref{tab:matrix} come from low-energy measurements that assume unitarity in the CKM matrix and no new loops in the relevant Feynman diagrams. This analysis will relax such assumptions and present different scenarios for interpretation of the provided results.

The signatures for $t$-channel processes involving $\vtb$, $\vtd$, and $\vts$ either in production or decay differ in three aspects: the number of reconstructed \PQb-tagged jets, the features of the jet involved in the reconstruction of the correct top quark candidate, and the kinematic features of the events as a result of different PDF contributions to production modes involving a \PQb, \PQs, or \PQd quark.
Henceforth, the $t$-channel process involving $\vtb$ in both production and decay will be referred to as $\STbb$, while $t$-channel processes involving $\vtb$ in only production or decay will be referred to as $\STbq$ and $\STqb$, respectively.
The signal channels and their corresponding cross sections times branching fractions from simulation are reported in Table~\ref{tab:abgfit}.
\begin{table*}[ht]
 \topcaption{For each of the production and decay vertices, the cross section times branching fraction for the corresponding signal process from simulation.  The uncertainties shown include those from the factorisation and renormalisation scales, the PDFs, and any experimental uncertainties, where appropriate.} 
  \label{tab:abgfit}  
\centering
\begin{tabular}{ lcc }
Production & Decay & Cross section $\times$ branching fraction (pb)\\
\hline
$\PQt\PW\PQb$ & $\PQt\PW\PQb$ & $ 217.0 \pm 8.4 $ \\
$\PQt\PW\PQb$ & ($\PQt\PW\PQs + \PQt\PW\PQd$) & $0.41 \pm 0.05 $ \\
$\PQt\PW\PQd$ & $\PQt\PW\PQb$ & $0.102 \pm 0.015 $ \\
$\PQt\PW\PQs$ & $\PQt\PW\PQb$ & $0.92 \pm 0.11 $ \\
\end{tabular}
\end{table*}
The cross sections are evaluated at NLO in the five-flavour scheme using \POWHEG for $\sigmatd$, $\sigmats$, and with \textsc{HATHOR}~\cite{hator} for $\sigmatb$. 

Multiple categories are defined in order to extract the contribution of the different $t$-channel processes, while at the same time discriminating against the background processes, mainly \ttbar and \wjets production. The majority of $t$-channel events populate categories with 2 or 3 jets, as defined above. The main backgrounds arise from \ttbar (all categories), \wjets (in the \twoJoneT and \threeJoneT categories), and QCD multijet (in the \twoJoneT category) processes. The signal processes taken into consideration give different contributions to the three categories, and it is possible to identify the most sensitive categories with respect to each process based on the respective signatures, as summarised in Table~\ref{tab:signalcategories}. The physics motivations leading to this strategy are described below.

Events from strong interaction \ttbar production, where one top quark decays through the $\PQt\PW\PQd$ or $\PQt\PW\PQs$ vertex ($\TTqb$), populate the \twoJoneT and \threeJoneT categories. Their small contribution to the \ttbar yield in such categories is covered by the \PQb-tagging uncertainty. Their signature in those categories is also found to be indistinguishable, within systematic uncertainties, from that of \ttbar when each top quark decays through the $\PQt\PW\PQb$ vertex and one \PQb jet does not pass either the kinematic or \PQb tagging requirements. 
For those reasons, all top quark decay modes of \ttbar pairs are treated as a single background source.

\begin{table*}[ht]
\topcaption{For each category, the corresponding signal process, the cross section times branching fraction expression, and the specific Feynman diagram from Fig.~\ref{fig:FG} are shown.} 
\label{tab:signalcategories}
\centering
\begin{tabular}{ lccc }
Category & Enriched in & Cross section $\times$ branching fraction & Feynman diagram\\
\hline
\twoJoneT & \STbb & $\sigmatb \BRtb$ & \ref{fig:FG}a\\
\threeJoneT & \STbq, \STqb & $\sigmatb \BRtq $, $\sigmatq \BRtb$ &  \ref{fig:FG}b, \ref{fig:FG}c, \ref{fig:FG}d\\
\threeJtwoT & \STbb &  $\sigmatb \BRtb$ &  \ref{fig:FG}a\\
\end{tabular}
\end{table*} 

The discrimination between the three signals $\STbq$, $\STqb$, and $\STbb$ is based on three characteristics.
First, for $\STbq$ events, only a single \PQb quark is present in the final state stemming from gluon splitting, thus resulting in a low-energy \PQb-tagged jet, while the jet coming from the top quark decay is usually not \PQb tagged. For $\STqb$ events, a single \PQb-tagged jet is produced in the top quark decay, and additional jets from gluon splitting are usually not \PQb tagged. Both $\STqb$ and $\STbq$ processes therefore differ from $\STbb$ by having a single \PQb quark in the final state, as opposed to two for the latter process. However, this feature can only be exploited when the jet from gluon splitting is energetic enough to be reconstructed. 
Second, further discrimination is achieved by exploiting the features of the reconstructed top quark candidates. The kinematic and angular properties of the decay products exhibit significant differences depending on whether the correct jet is chosen, or if the jet that originated from the quark produced in the gluon splitting is used. For $\STbq$ events, the top quark reconstructed with the correct jet assignment usually does not use the \PQb-tagged jet in the event, while for $\STbb$ and $\STqb$, the top quark candidate is reconstructed by using the \PQb-tagged jet in the majority of cases. It is therefore possible to differentiate between the $\STbb$ and $\STbq$ processes by comparing the features of top quark candidates reconstructed with or without \PQb-tagged jets. 
Finally, different PDFs are involved in $\STbb$ and $\STqb$ processes, the latter drawing contributions from valence \PQd quarks as well. Therefore, the kinematic properties of final-state particles may differ from the other channels.
The second characteristic, related to the correctness of the top quark reconstruction hypothesis, proves to be the strongest amongst the three mentioned criteria. While the $\STbq$ and the $\STbb$ processes can be differentiated by using this characteristic, the $\STqb$ and the $\STbb$ productions cannot, because their final-state signatures exhibit the same features.

The \twoJoneT category is populated by events that depend on $\vtb$ in both production and decay, where the single reconstructed \PQb jet comes in the majority of cases (85\%) from top quark decays, and for the remaining cases from the second \PQb jet from gluon splitting. This means that the jet from the second \PQb quark fails either the jet $\pt$ requirement or the \PQb tag requirement, or both. Events coming from a process for which $\vtd$ or $\vts$ are involved, either in production or in decay, populate this category as well, with either the \PQb-tagged jet coming from top quark decay or the secondary \PQb quark from gluon splitting. 

For $t$-channel signal events from all four processes in Fig.~\ref{fig:FG}, the most distinctive features that allow the discrimination against backgrounds in the \twoJoneT category rely on the fact that the second jet stems from the recoiling quark. For this reason the non-\PQb-tagged jet is not used for the top quark reconstruction. This category is the one where the highest discrimination power for $\STbb$ against backgrounds is achieved by making use of the features of the top quark decay products, such as the reconstructed top quark mass and \mtw, and of the recoiling jet. However, the discrimination power with respect to other $t$-channel mechanisms is poor since jets from gluon splitting are typically not energetic enough to pass the \pt threshold, making it impossible to reconstruct two different top quark candidates.

The \threeJoneT category is also populated by all $t$-channel processes of interest, but it differs from \twoJoneT in the fact that it accommodates events in which the jet from gluon splitting has a higher \pt on average. For both the \twoJoneT and \threeJoneT categories, when the top quark decays through $\PQt\PW\PQd$,\PQs vertices, the jet coming from the top quark usually does not pass the \PQb tagging requirement since it stems from the hadronisation of a light quark. In all other cases, this jet passes the \PQb tagging requirement, given the efficiency of the tagging algorithm. 

The \threeJoneT category is enriched in $t$-channel events by requiring $\abs{\etalj}>2.5$, where $\etalj$ is the pseudorapidity of the most forward jet.
The two jets other than the most forward one are used to reconstruct the two top quark candidates. If the event is from the $\STbq$ process, the \PQb-tagged jet in the \threeJoneT category will stem from gluon splitting, and the additional jet will have a higher chance of being the one coming from the top quark decay to an \PQs or \PQd quark. Variables of interest in this case are constructed by making use of the \PQb jet and the least forward jet of the remaining two, referred to as the extra jet. Such variables include the invariant mass of the lepton plus jet system (either the \PQb jet or the extra jet), and several top quark kinematic variables constructed using a combination of the extra jet, the lepton, and \ptmiss. 

\begin{figure*}[!ht]
\centering
\includegraphics[width=0.45\textwidth]{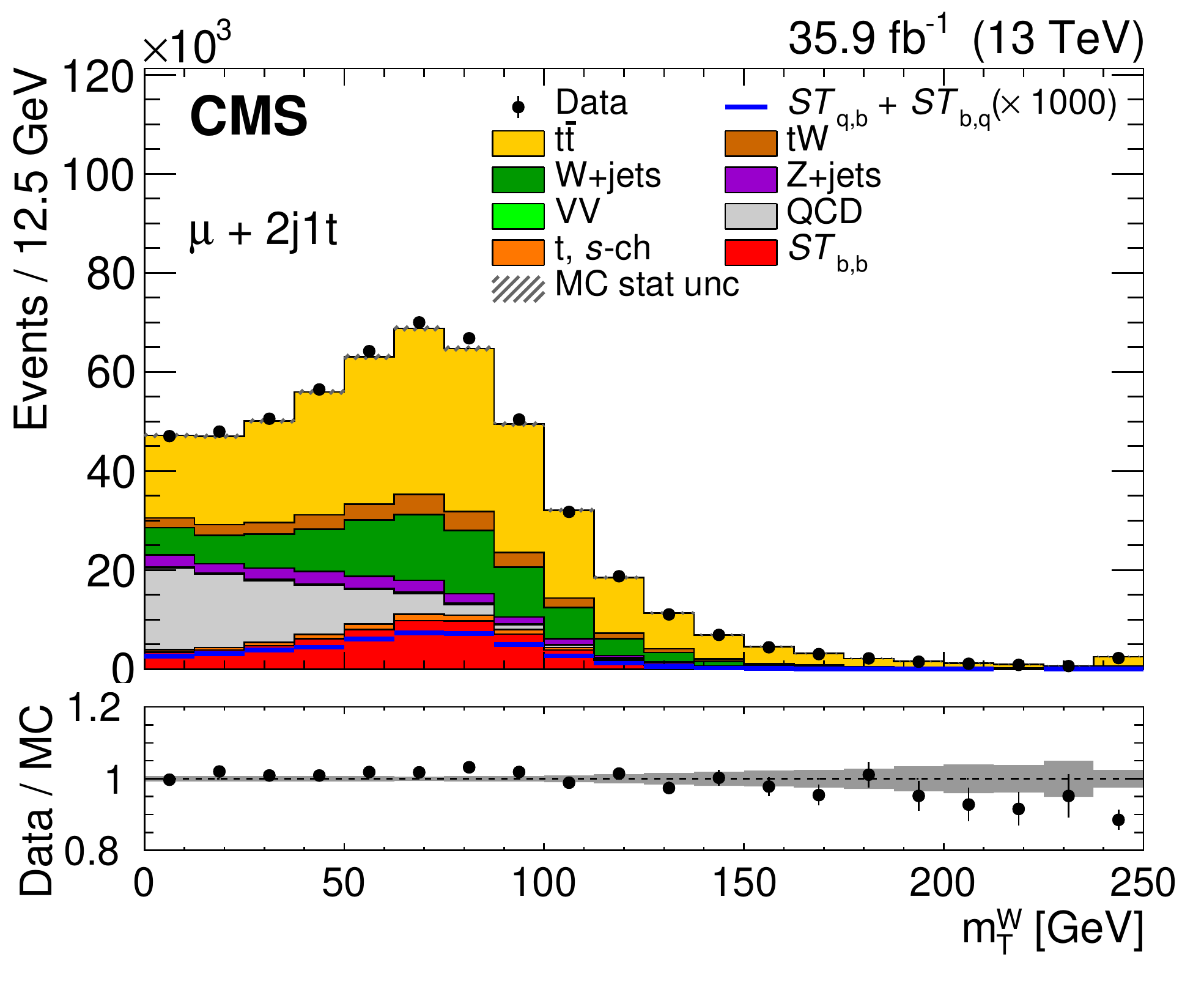}
\hspace{0.05\textwidth}
\includegraphics[width=0.45\textwidth]{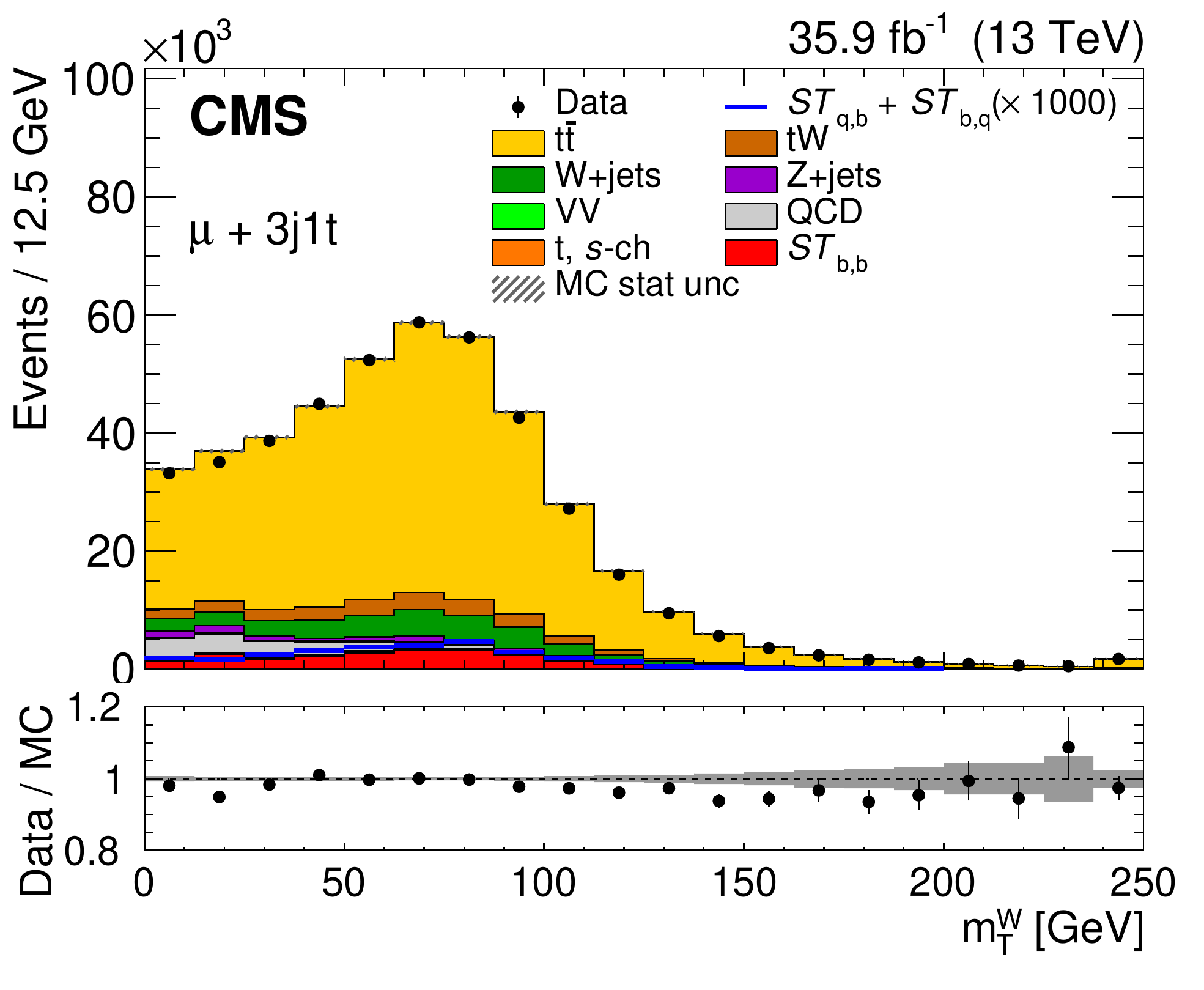}
\includegraphics[width=0.45\textwidth]{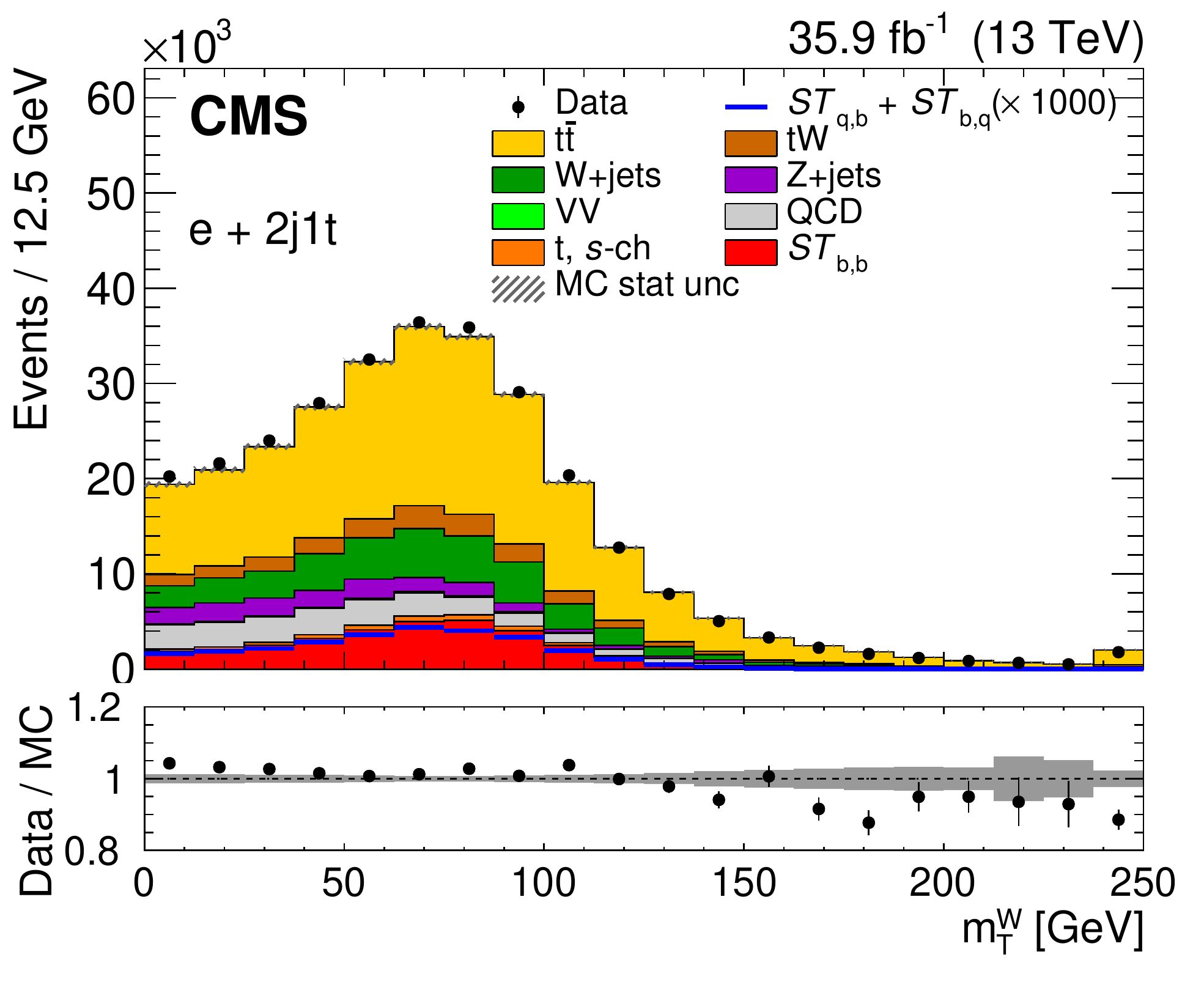}
\hspace{0.05\textwidth}
\includegraphics[width=0.45\textwidth]{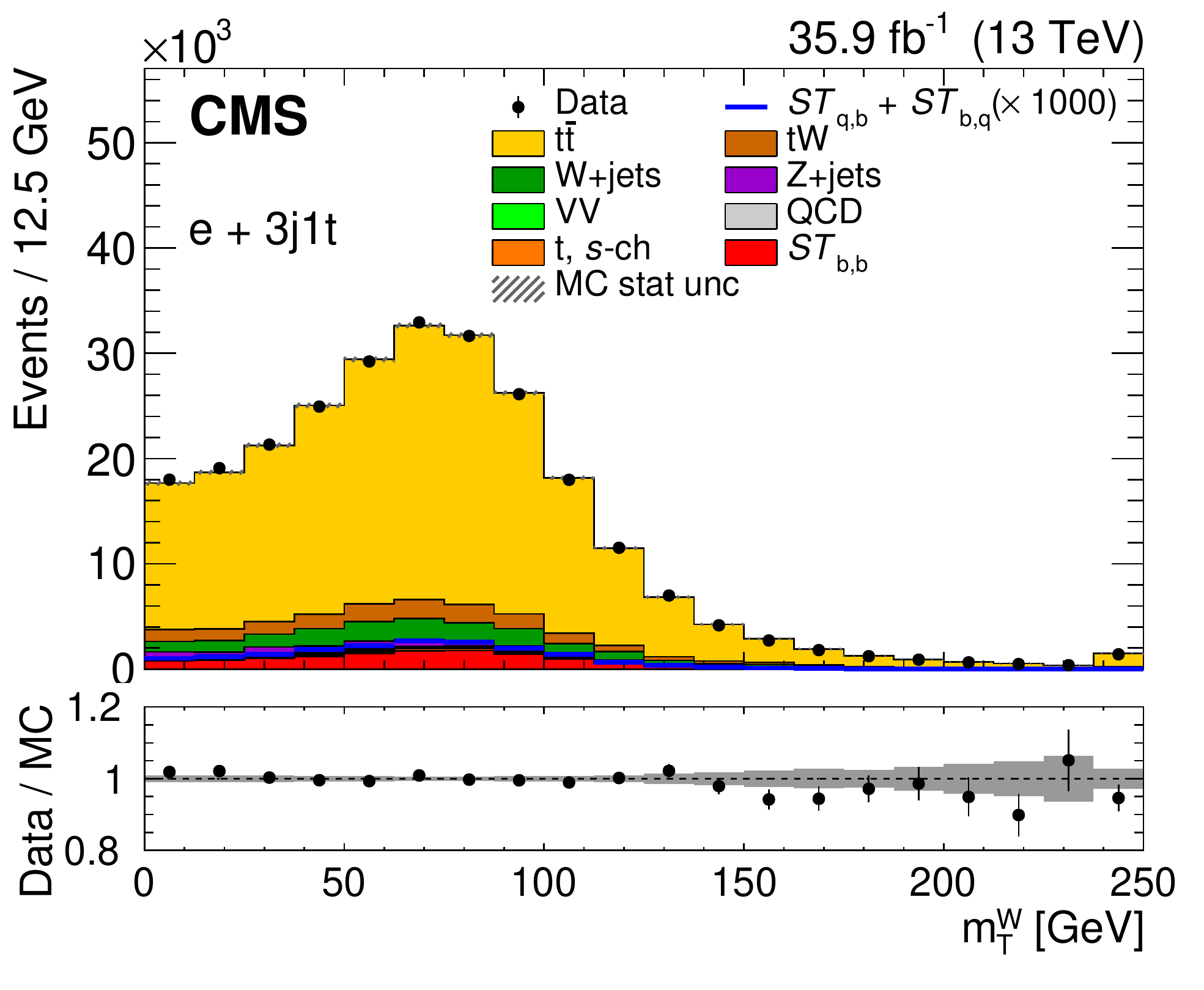}
\caption{\label{fig:nj1t_mtw} The \mtw distribution from data (points) and simulation (shaded histograms) in the \twoJoneT (left)  and \threeJoneT (right) categories for the muon (upper) and electron  (lower) channels. The vertical lines on the points and the hatched bands show the experimental and MC statistical uncertainties, respectively. The expected distribution from the $\STqb + \STbq$ processes (multiplied by a factor of 1000) is shown by the solid blue line. The lower panels show the ratio of the data to the MC prediction.}
\end{figure*}

In both the \twoJoneT and the \threeJoneT categories, \mtw is also used to discriminate between the QCD multijet background and other processes. An event category depleted of QCD multijet background is defined by adding the requirement $\mtw>50\GeV$. Figure~\ref{fig:nj1t_mtw} shows the $\mtw$ distribution from data and simulations in the \twoJoneT and \threeJoneT categories for the muon (upper plots) and electron (lower plots) channels, where the QCD multijet background is normalised to the result of the fit. 

In the \threeJtwoT category, there are two \PQb jets, one produced from the top quark decay and another from gluon splitting. Both \PQb jets are used to reconstruct a top quark candidate and its corresponding variables. In this case, the $\mtw>50\GeV$ requirement is unnecessary since the $\QCD$ contamination is negligible and the dominant background process is $\ttbar$. No requirement on $\etalj$ is needed either since the category is dominated by the $\STbb$ process and the combinatorial top quark background is small.

Multivariate analyses are then performed by using boosted decision trees (BDT) in order to obtain appropriate discriminating variables, henceforth referred to as BDT discriminators, in the three categories, for both muons and electrons. 
The processes used as signal or background in the training are the following:
\begin{itemize}
\item In the \twoJoneT category, the single top quark $\STbb$ process is considered as signal and \ttbar and \wjets processes as background.
\item In the \threeJoneT category, the single top quark $\STqb$ process is considered as signal and the $\STbb$, \ttbar, and \wjets processes as background.
\item In the \threeJtwoT category, the single top quark $\STbb$ process is considered as signal and \ttbar as background.
\end{itemize}

The variables used in the \twoJoneT category training are: the \abseta of the non-\PQb-tagged jet, the reconstructed top quark mass, the cosine of the angle between the \PW boson momentum in the top quark rest frame and the momentum of the lepton in the \PW boson rest frame, the cosine of the polarisation angle defined as the angle between the direction of the lepton and the light-quark momenta in the top quark rest frame, the invariant mass of the lepton and \PQb-tagged jet system, and the invariant mass of the lepton and forward jet system.

The variables used in the \threeJoneT category training are: the \abseta of the most forward non-\PQb-tagged jet, the mass of the top quark when it is reconstructed with the \PQb-tagged jet (\PQb-top quark), the cosine of the angle between the \PW boson momentum in the \PQb-top quark rest frame and the momentum of the lepton in the \PW boson rest frame, the cosine of the polarisation angle defined as the angle between the direction of the lepton and the light-quark momenta in the \PQb-top quark rest frame, \ptmiss, \mtw, the invariant mass of the lepton and \PQb-tagged jet system, the invariant mass of the lepton and extra jet system, the invariant mass of the lepton and forward jet system, the number of low-\pt jets, the mass of the top quark when it is reconstructed with the non-\PQb-tagged jet (non-\PQb-top quark), the cosine of the angle between the \PW boson momentum in the non-\PQb-top quark rest frame and the momentum of the lepton in the \PW boson rest frame, the cosine of the polarisation angle defined as the angle between the direction of the lepton and the light-quark momenta in the non-\PQb-top quark rest frame, and the value of the MVA \PQb tagger discriminator when applied to the non-\PQb-tagged jet.

The variables used in the \threeJtwoT category training are: the \abseta of the non-\PQb-tagged jet, the mass of the top quark when it is reconstructed with the highest-\pt \PQb-tagged jet (leading top quark), the cosine of the angle between the \PW boson momentum in the leading top quark rest frame and the momentum of the lepton in the \PW boson rest frame, the cosine of the polarisation angle defined as the angle between the direction of the lepton and the light-quark momenta in the leading top quark rest frame, \ptmiss, \mtw, the invariant mass of the lepton and the highest-\pt \PQb-tagged jet system, the invariant mass of the lepton and lower-\pt \PQb-tagged jet system, the invariant mass of the lepton and light-jet system, the number of low-\pt jets, the mass of the top quark when it is reconstructed with the lower-\pt \PQb-tagged jet (non-leading top quark), the cosine of the angle between the \PW boson momentum in the non-leading top quark rest frame and the momentum of the lepton in the \PW boson rest frame, the cosine of the polarisation angle defined as the angle between the direction of the lepton and the light-quark momenta in the non-leading top quark rest frame, and the difference in $\eta$ between the two \PQb-tagged jets.

\begin{figure*}[!ht]
\centering
\includegraphics[width=0.45\textwidth]{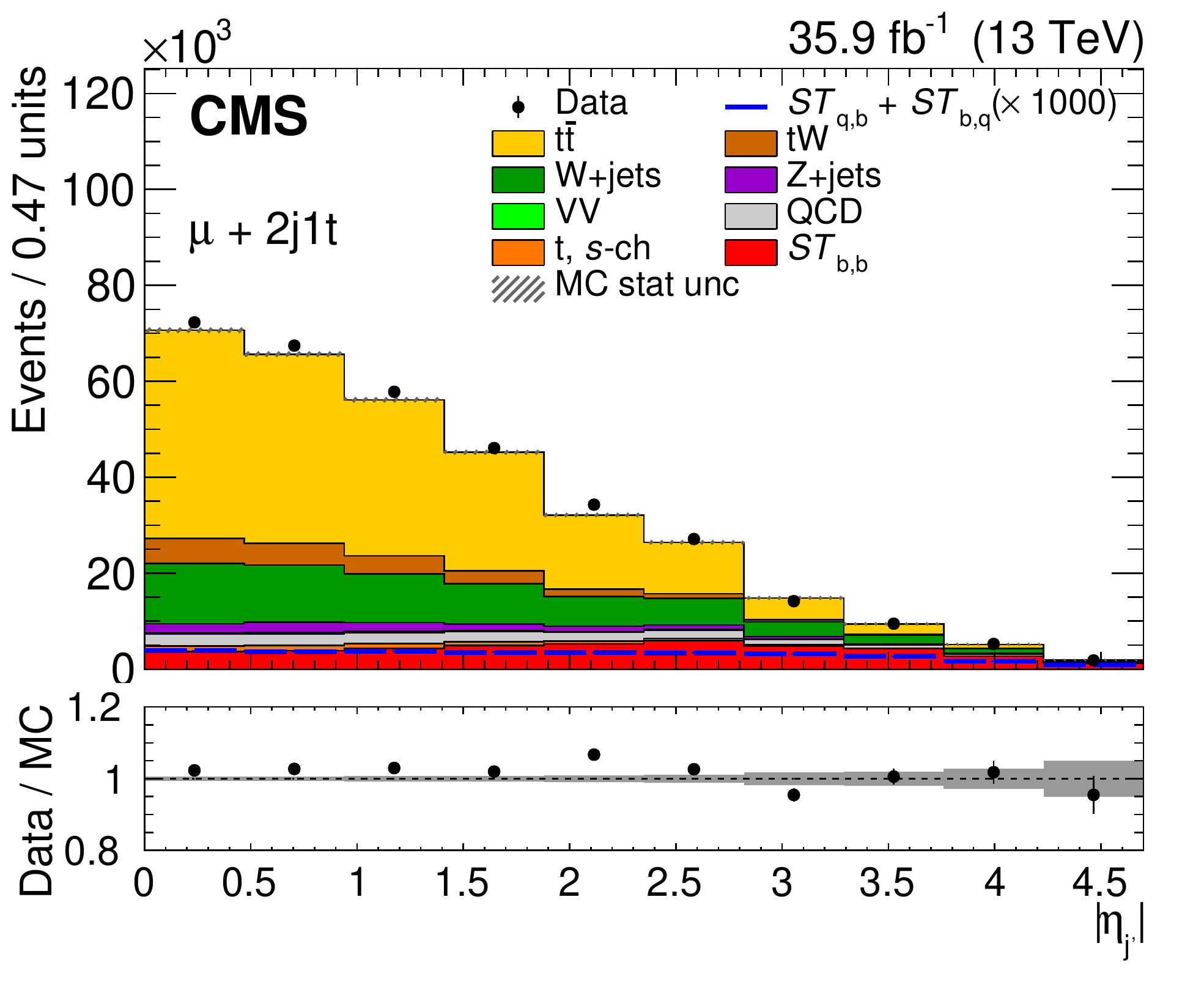}
\hspace{0.05\textwidth}
\includegraphics[width=0.45\textwidth]{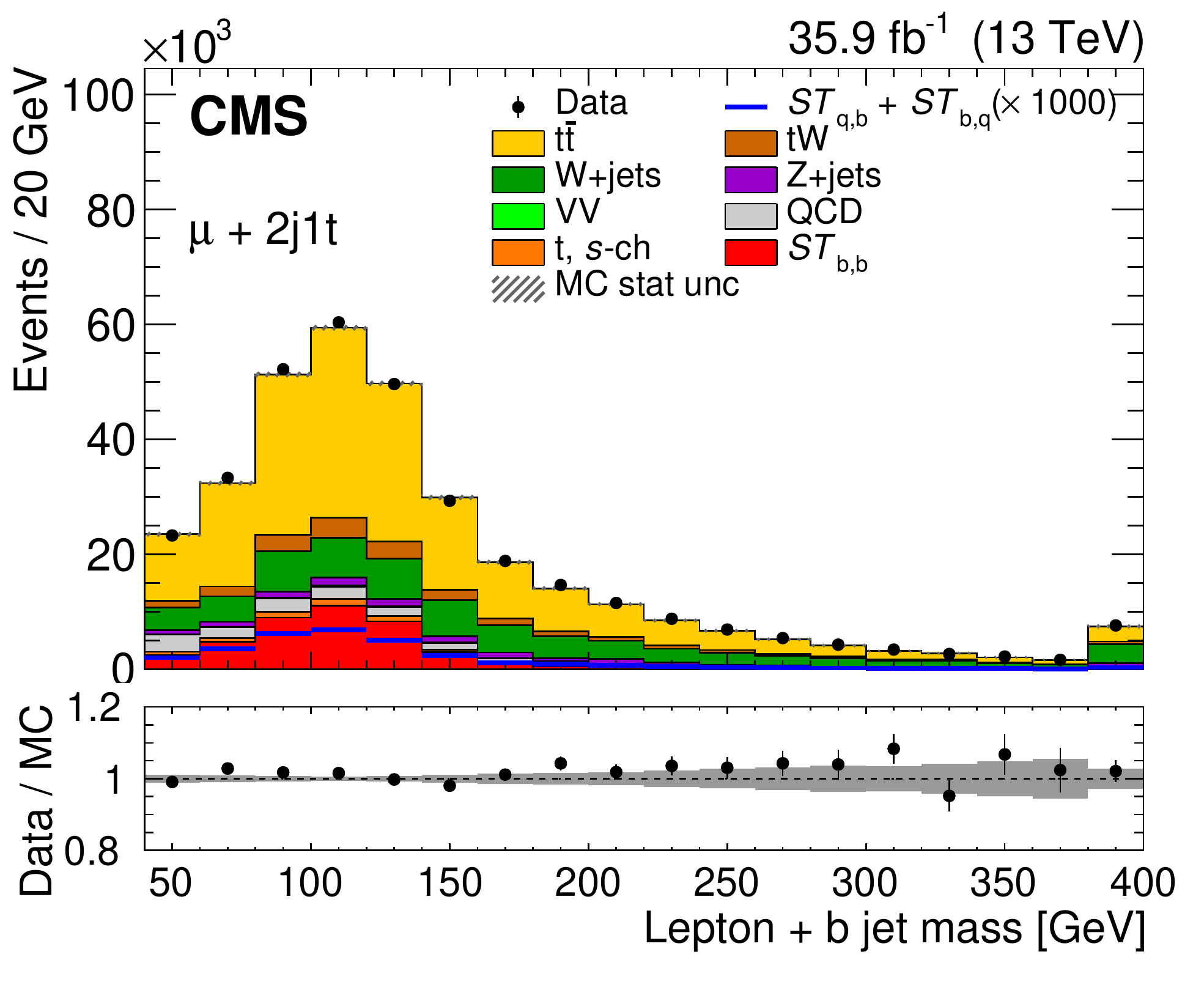}
\includegraphics[width=0.45\textwidth]{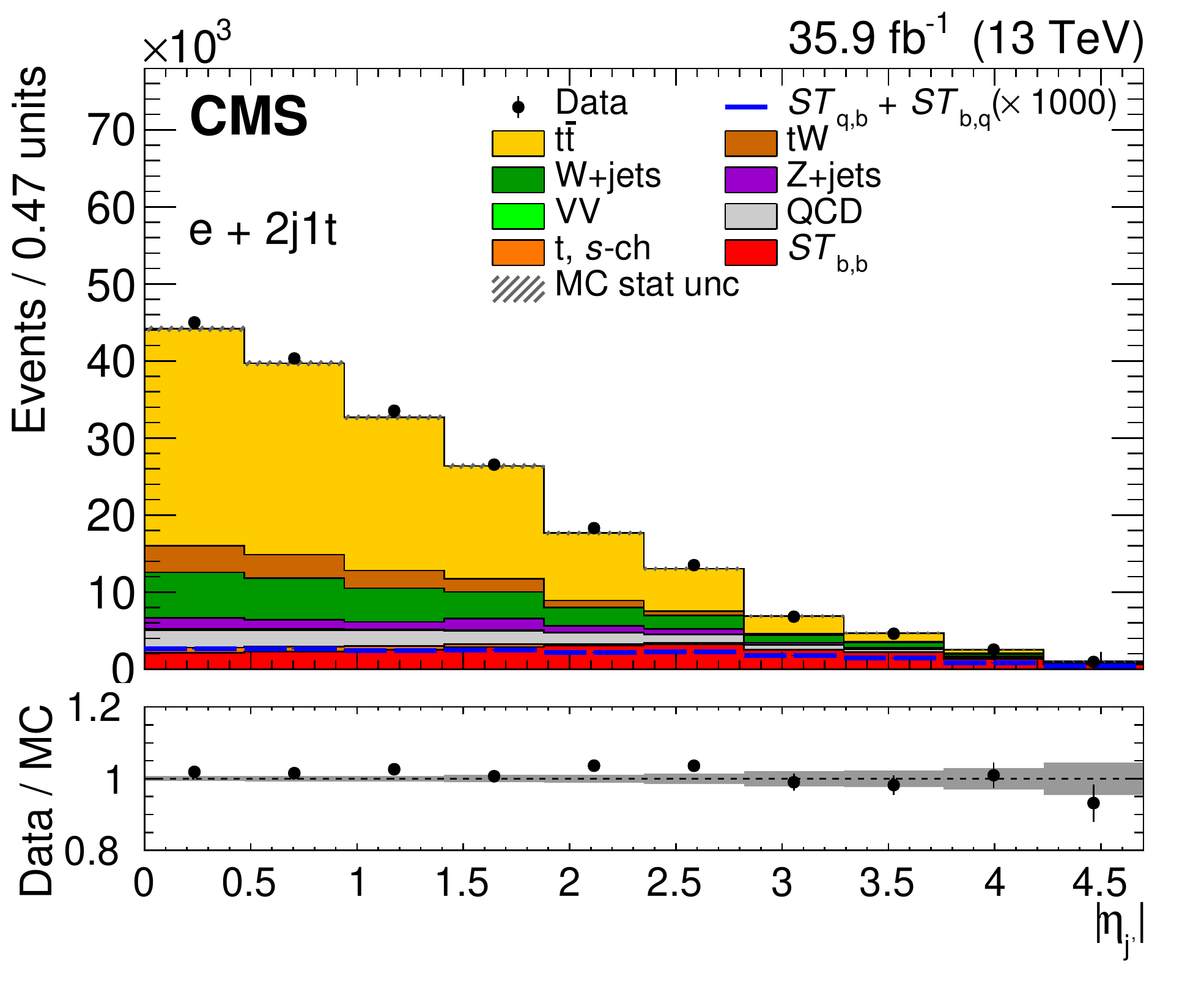}
\hspace{0.05\textwidth}
\includegraphics[width=0.45\textwidth]{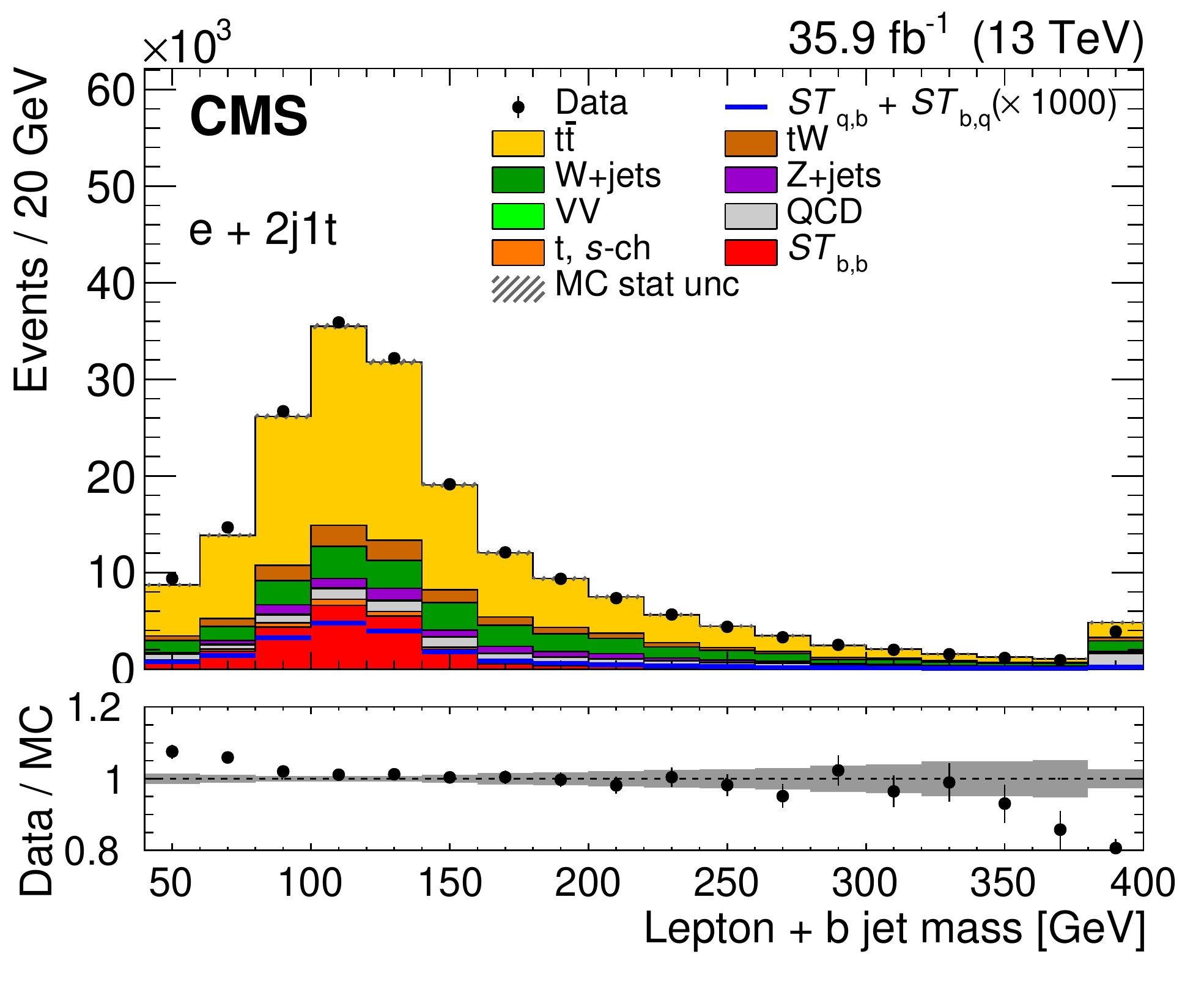}
\caption{\label{fig:2j1t_variables} Distributions of the two most discriminating variables from data (points) and simulation (shaded histograms) in the \twoJoneT category: the \abseta of the non-\PQb-tagged jet \etalj (left) and the invariant mass of lepton and \PQb jet momenta system (right), shown for the muon (upper) and electron (lower) channels, respectively. The vertical lines on the points and the hatched bands show the experimental and MC statistical uncertainties, respectively.  The expected distribution from the $\STqb + \STbq$ processes (multiplied by a factor of 1000) is shown by the solid blue line. The lower panels show the ratio of the data to the MC prediction.}
\end{figure*}

\begin{figure*}[!ht]
\centering
\includegraphics[width=0.45\textwidth]{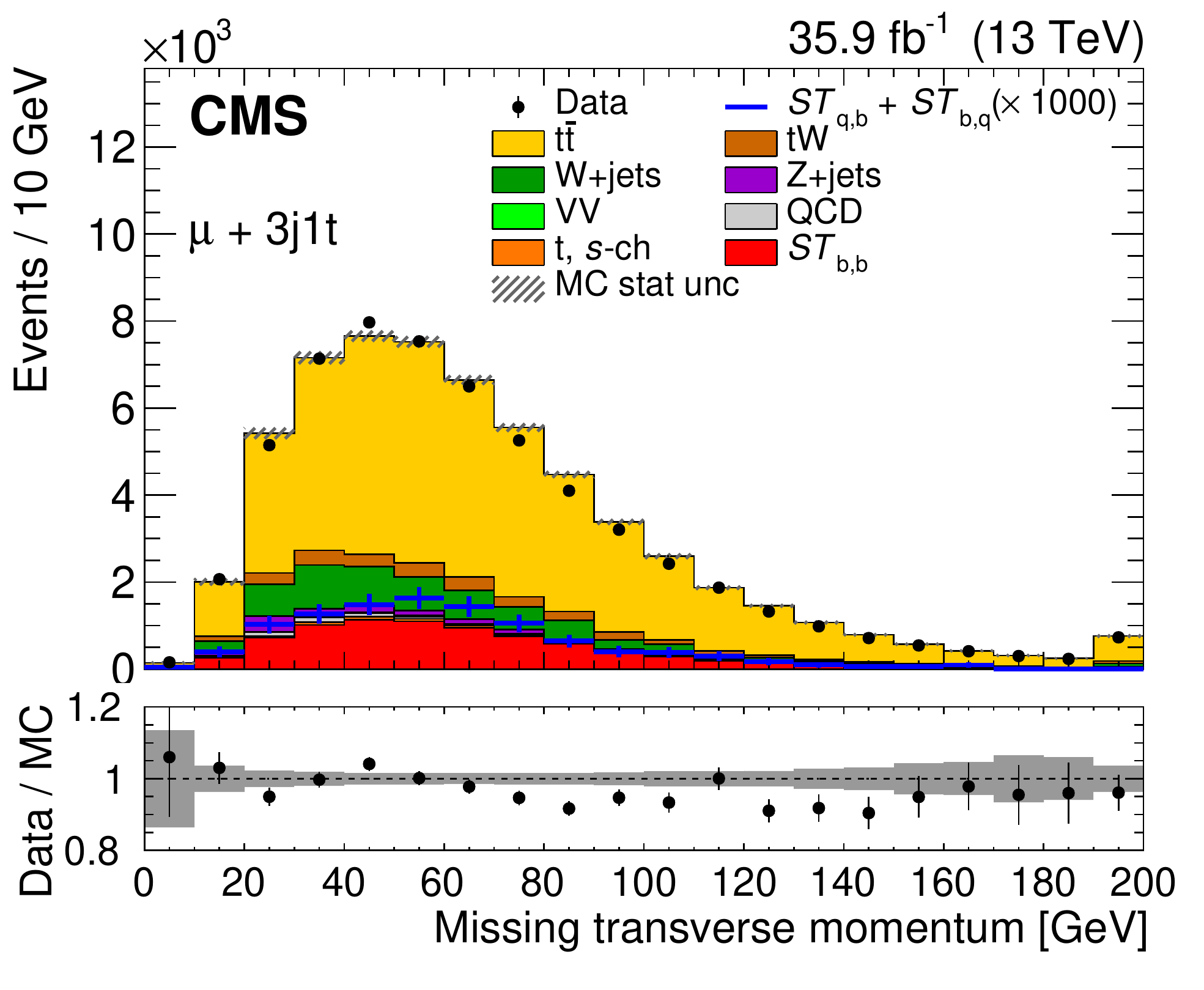}
\hspace{0.05\textwidth}
\includegraphics[width=0.45\textwidth]{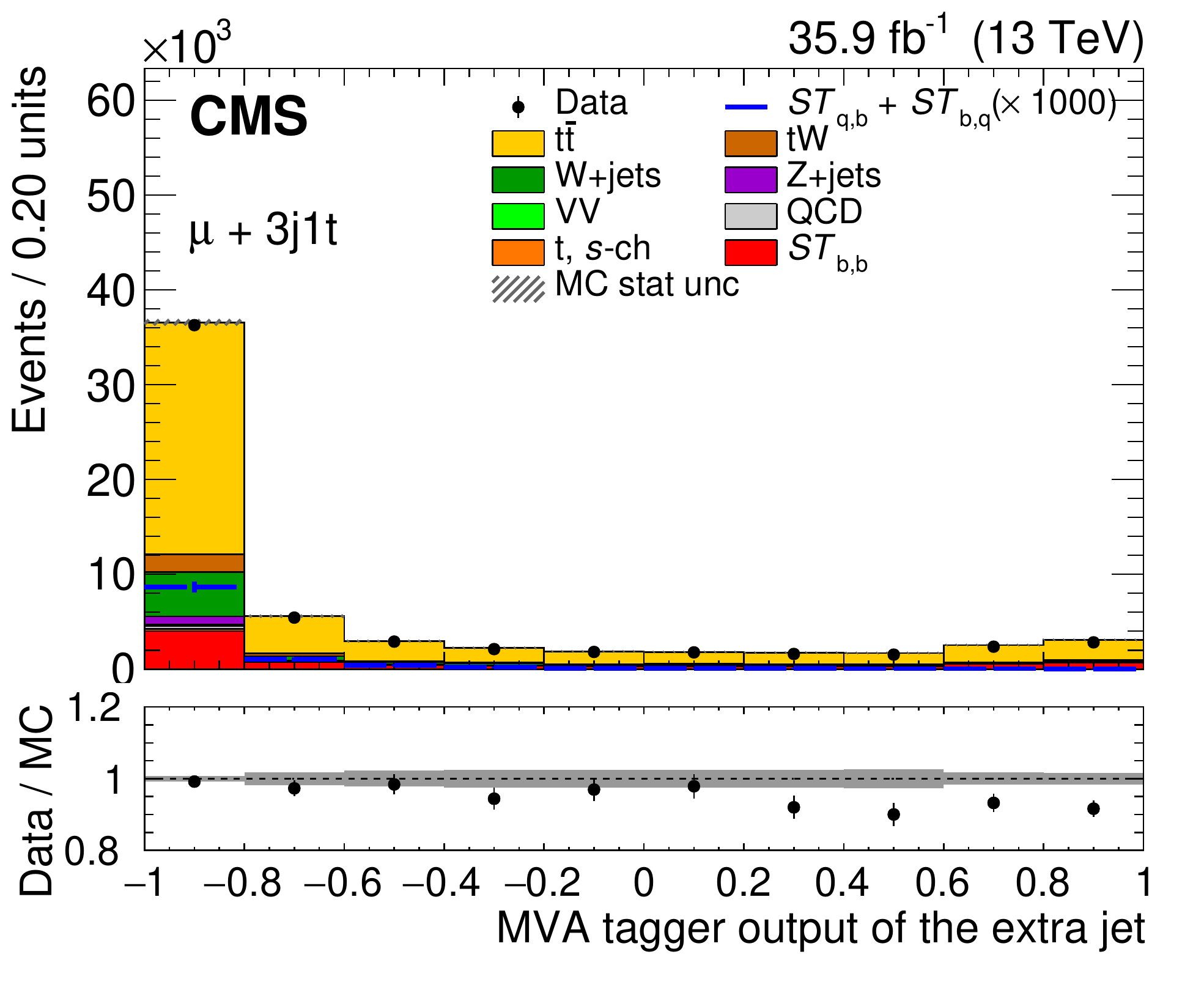}
\includegraphics[width=0.45\textwidth]{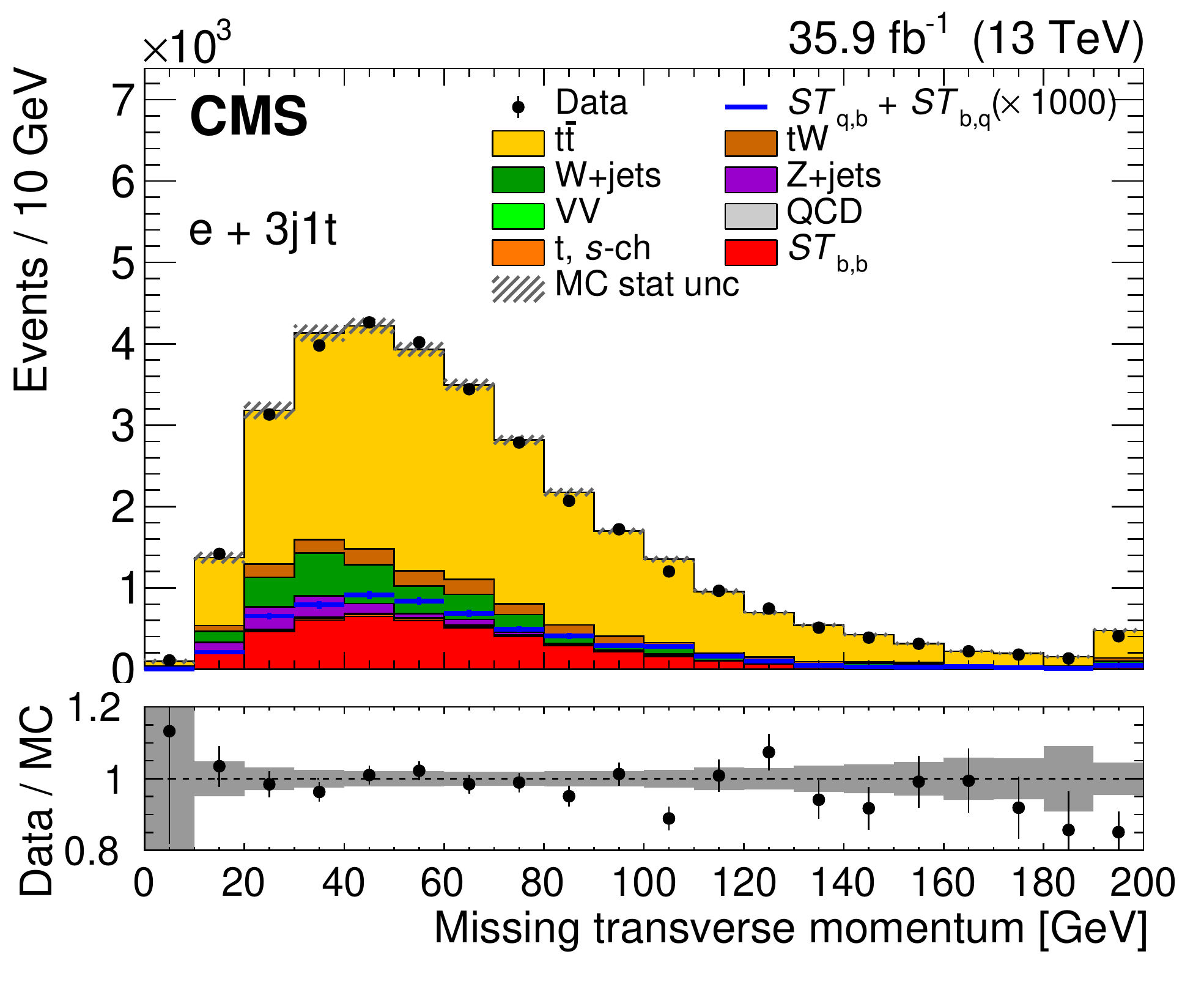}
\hspace{0.05\textwidth}
\includegraphics[width=0.45\textwidth]{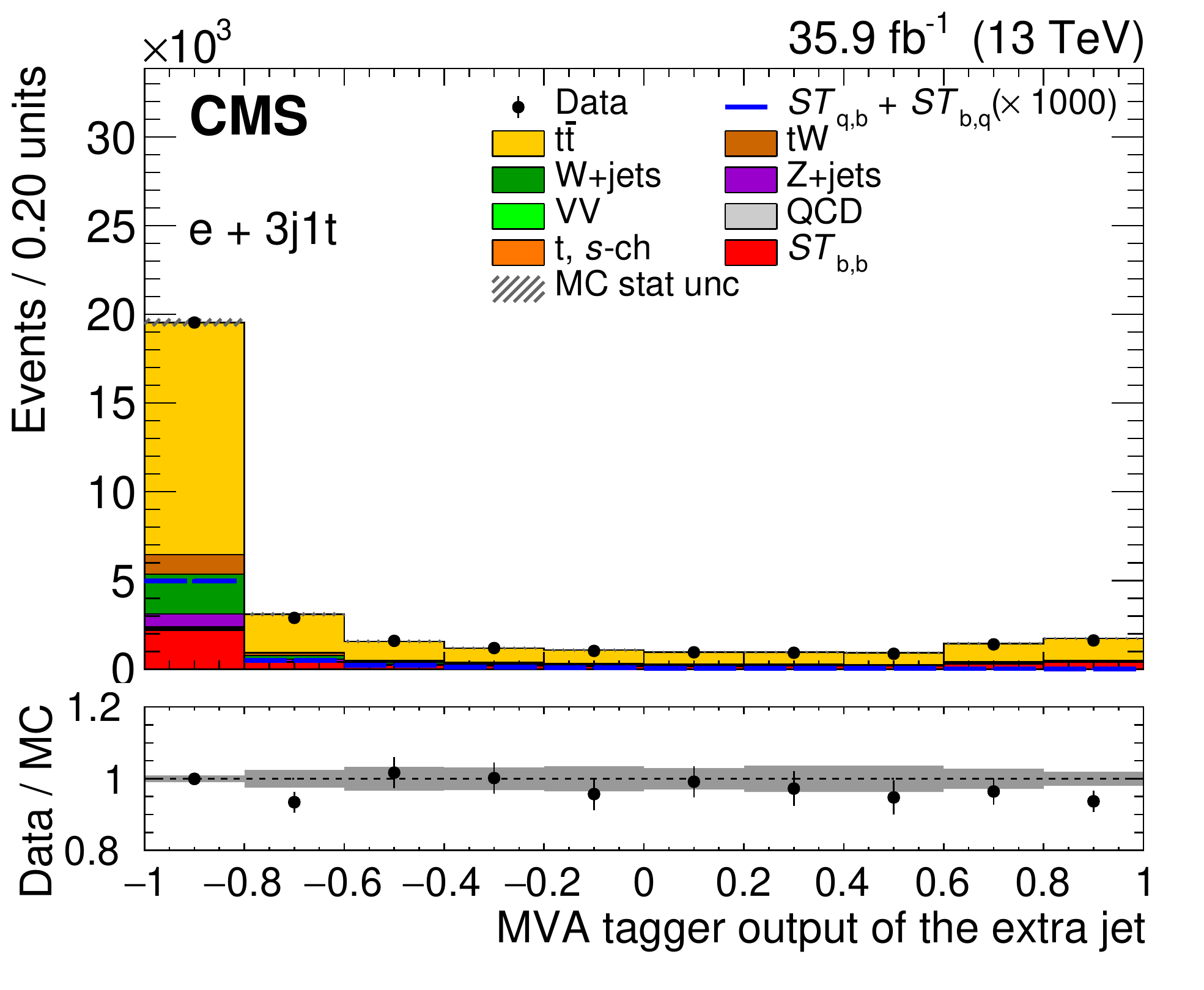}
\caption{\label{fig:3j1t_variables} Distributions of the two most discriminating variables from data (points) and simulation (shaded histograms) in the \threeJoneT category: the \ptmiss in the transverse plane (left) and the value of the MVA \PQb tagger discriminator when applied to the extra jet (right) are shown for the muon (upper) and electron (lower) channels, respectively. The vertical lines on the points and the hatched bands show the experimental and MC statistical uncertainties, respectively.  The expected distribution from the $\STqb + \STbq$ processes (multiplied by a factor of 1000) is shown by the solid blue line. The lower panels show the ratio of the data to the MC prediction.}
\end{figure*}

\begin{figure*}[!ht]
\centering
\includegraphics[width=0.45\textwidth]{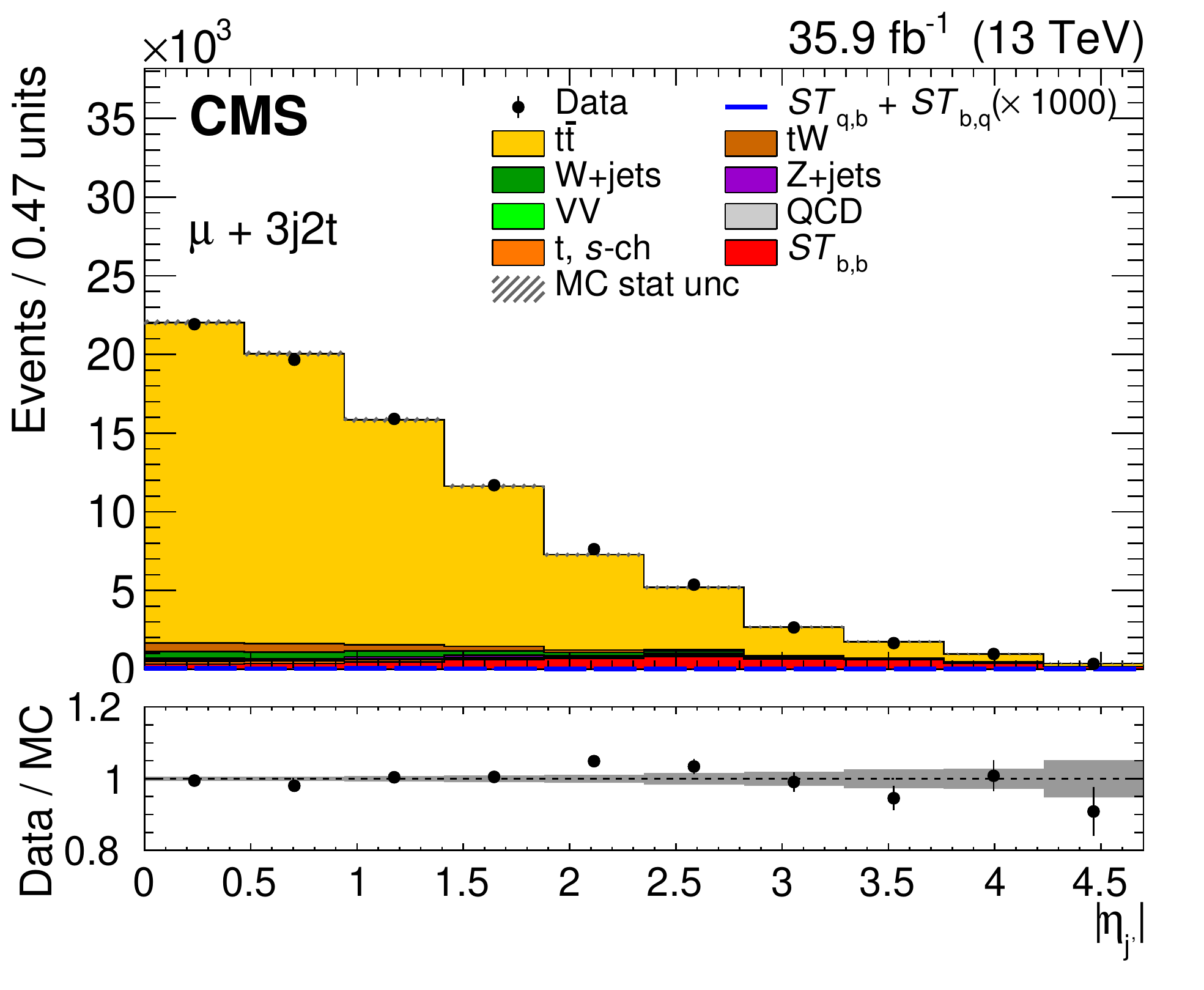}
\hspace{0.05\textwidth}
\includegraphics[width=0.45\textwidth]{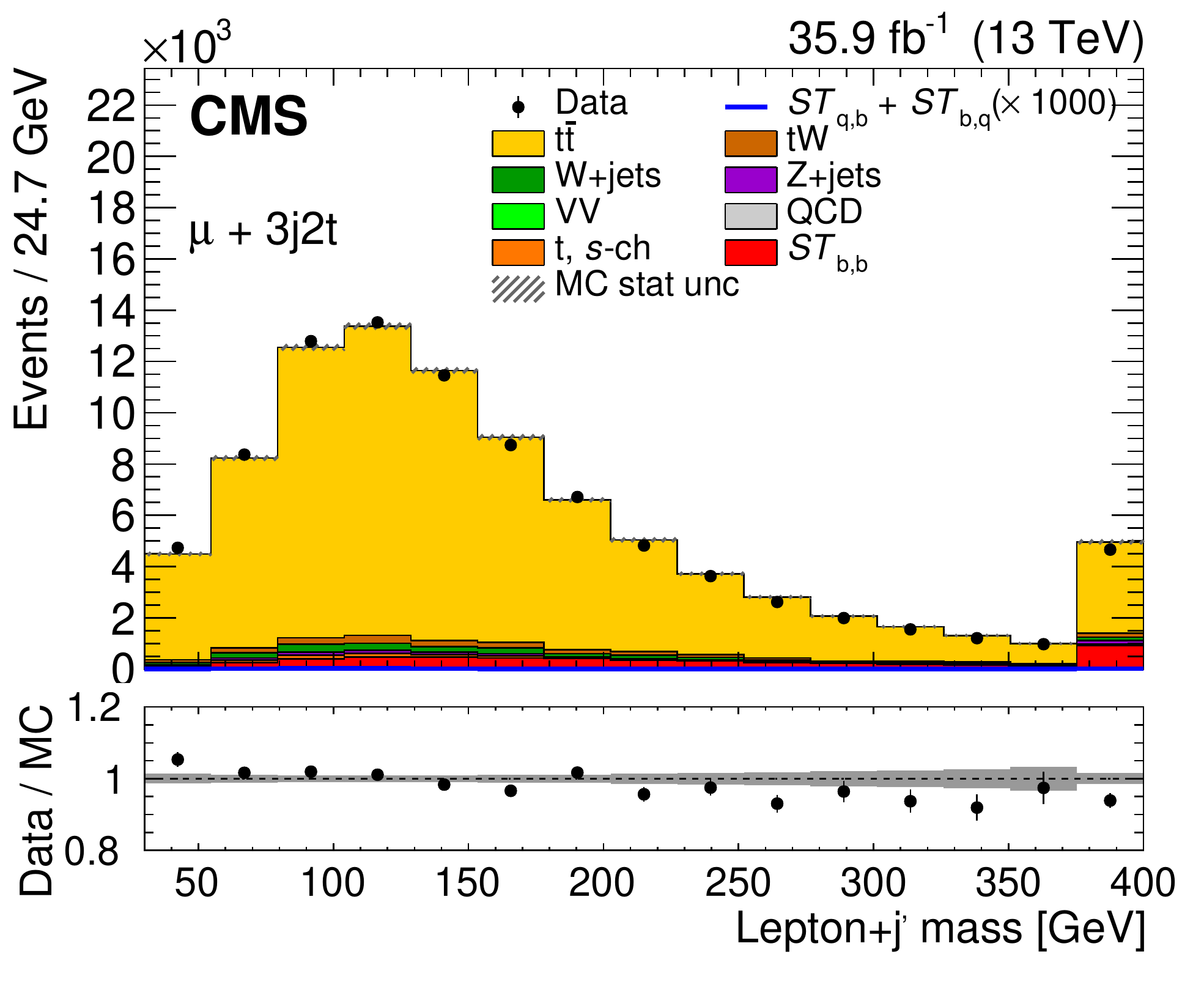}
\includegraphics[width=0.45\textwidth]{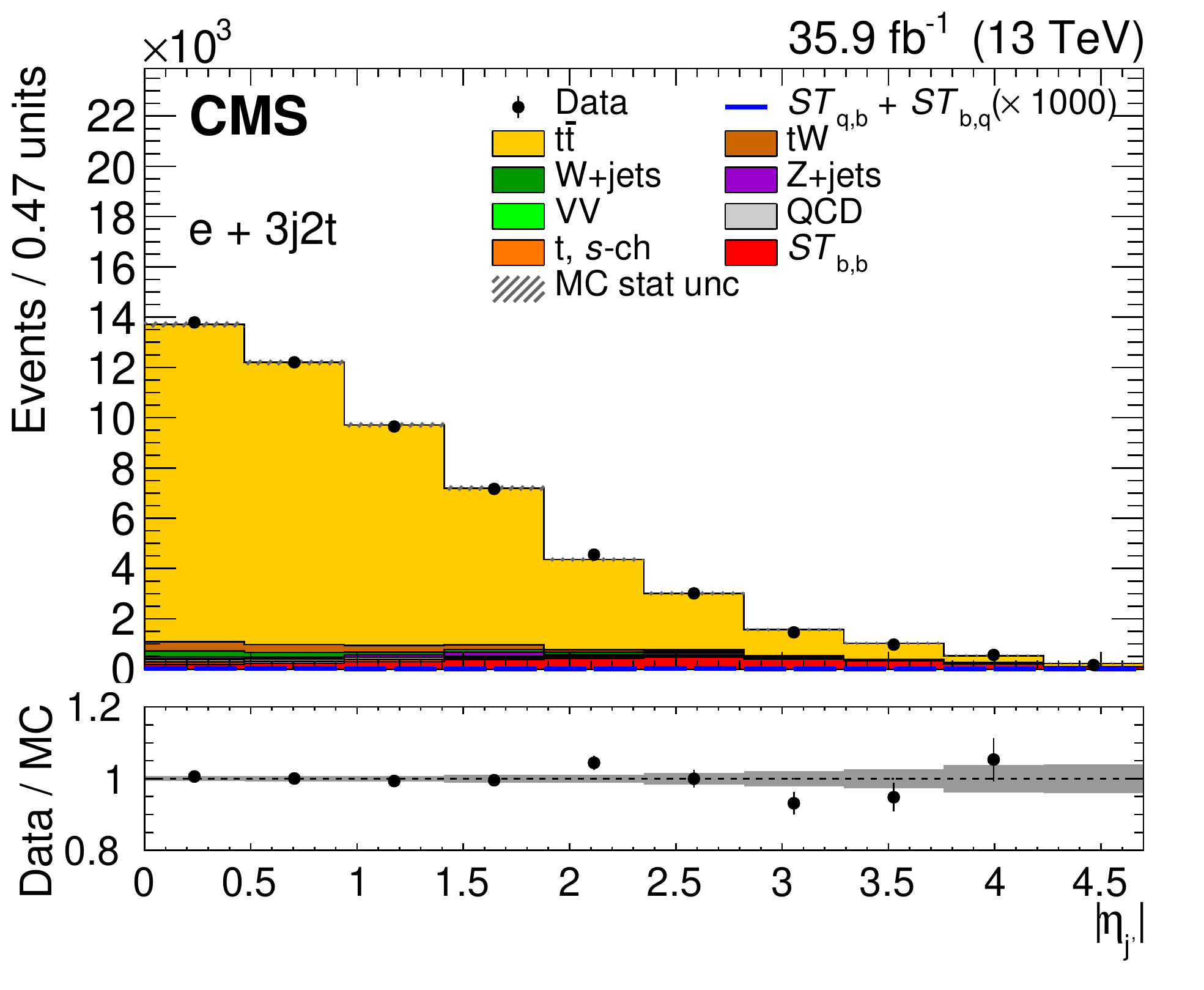}
\hspace{0.05\textwidth}
\includegraphics[width=0.45\textwidth]{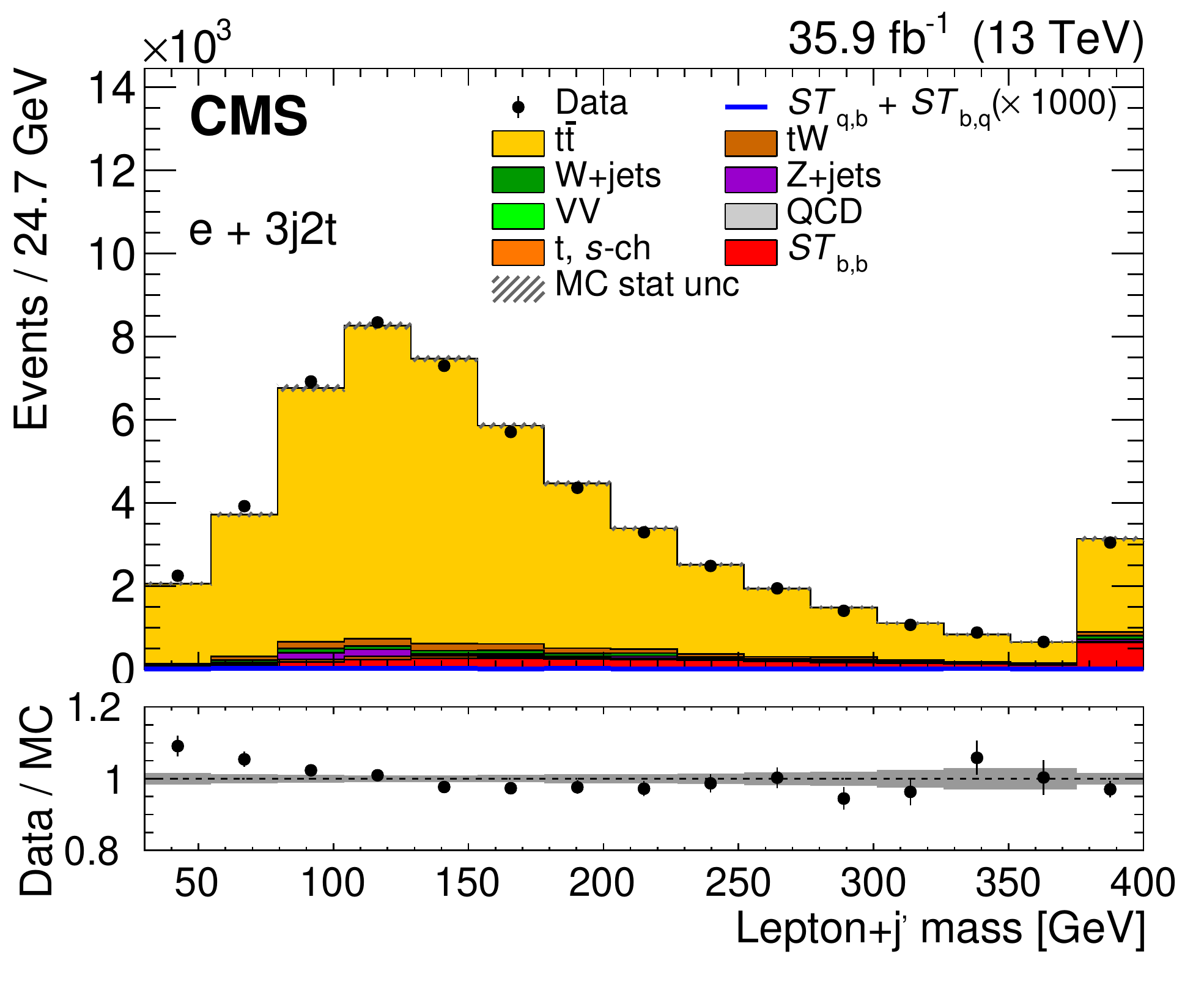}
\caption{\label{fig:3j2t_variables} Distributions of the two most discriminating variables from data (points) and simulation (shaded histograms) in the \threeJtwoT category:  the \abseta of the non-\PQb-tagged jet \etalj (left) and the invariant mass of lepton and non-\PQb-tagged jet system (right) are shown for the muon (upper) and electron (lower) channels, respectively. The vertical lines on the points and the hatched bands show the experimental and MC statistical uncertainties, respectively.  The expected distribution from the $\STqb + \STbq$ processes (multiplied by a factor of 1000) is shown by the solid blue line. The lower panels show the ratio of the data to the MC prediction.}
\end{figure*}

Figures \ref{fig:2j1t_variables}--\ref{fig:3j2t_variables} show the distributions of the most discriminating variables in the \twoJoneT, \threeJoneT, and \threeJtwoT categories, respectively.

\section{Systematic uncertainties}
\label{sec:systs}
Several sources of systematic uncertainties are considered in the analysis, divided in two groups depending on the treatment: uncertainties labelled as ``profiled'' are treated as nuisance parameters and profiled in the fit procedure described in Section~\ref{sec:ckmfit}, while those labelled as ``nonprofiled''  are estimated as the difference between the result of the fit procedure by varying the systematic scenario.
These latter uncertainties include the sources related to the modelling of the signal process, which cannot be constrained from the measurement since they apply to the full phase space and not only to the region in which the measurement is performed. Also included are the jet energy scale and resolution uncertainties, which play a major role in events featuring hadronic activity in the high-pseudorapidity region of the detector. They are also intertwined with the uncertainties in the modelling of the hadronisation and cause a larger uncertainty in the signal acceptance, which was not the case for previous measurements~\cite{Sirunyan:2018rlu,Sirunyan:2019hqb}. For these reasons, a more conservative approach is preferred and these uncertainties are not profiled in the fit.

The impact of nonprofiled uncertainties is determined by repeating the analysis using varied templates according to the systematic uncertainty sources under study in the fit, instead of the nominal templates. The uncertainty due to a certain source is then taken as half the difference between the results for up and down variations of the effect. 
In the following, the different uncertainty sources that are considered in the analysis are briefly described. For the sake of simplicity and better readability, they are grouped into profiled and nonprofiled uncertainties.

Profiled uncertainties
\begin{itemize}
\item \textit{Limited size of simulated event samples}: The statistical uncertainty due to the limited size of the simulated event samples is evaluated for each bin with the Barlow--Beeston ``light'' method \cite{Barlow:1993dm, bb_light}. 
\item \textit{Lepton trigger and reconstruction}: Single-muon and single-electron trigger and reconstruction efficiencies are estimated with a ``tag-and-probe'' method~\cite{Khachatryan:2010xn} from Drell--Yan events with the dilepton invariant mass in the \PZ boson peak.
\item \textit{Pileup}: The uncertainty in the average expected number of pileup interactions is propagated as a source of systematic uncertainty by varying the total $\Pp\Pp$ inelastic cross section by $\pm 4.6\%$~\cite{Sirunyan:2018nqx}.
\item \textit{\ttbar modelling}: The following uncertainty sources cover potential mismodelling of the \ttbar process. Their effect is considered on both the acceptance and the cross section.
\begin{itemize}
\item \textit{\ttbar renormalisation and factorisation scale uncertainties ($\mu_{\mathrm{R}}/\mu_{\mathrm{F}}$)}: The uncertainties caused by variations in the renormalisation and factorisation scales are considered by reweighting the BDT response distributions with different combinations of doubled/halved renormalisation and factorisation scales with respect to the nominal value of 172.5\GeV.
\item \textit{Matching of matrix element and parton shower (ME-PS matching)}: The parameter that controls the matching between the matrix element level calculation and the parton shower, and that regulates the high-\pt radiation in the simulation is varied within its uncertainties.
\item \textit{Initial- and final-state radiation:} The impact of variations in the initial-state and final-state radiation is studied by comparing the nominal sample with dedicated \ttbar samples.
\item \textit{Underlying event:} The effect of uncertainties in the modelling of the underlying event is studied by comparing the nominal sample with dedicated \ttbar samples.
\end{itemize}
\item \textit{QCD multijet background process normalisation}: The QCD multijet background yield is assigned a $50\%$ uncertainty, which is chosen conservatively to be much larger than the uncertainty from the \mtw fit.
\item \textit{\wjets composition}: A separate uncertainty is dedicated to the fraction of \wjets events where the forward jet is generated by the parton showering.
\item \textit{Other backgrounds $\mu_{\mathrm{R}}/\mu_{\mathrm{F}}$}: In addition to \ttbar, the uncertainties due to variations in the renormalisation and factorisation scales are studied for the $\PQt\PW$ and \wjets processes by reweighting the distributions with weights corresponding to different combinations of halved or doubled factorisation and renormalisation scales. The effect is estimated for each process separately.
\item \textit{PDF for background processes}: The uncertainty due to the choice of PDF is estimated using reweighted histograms derived from all PDF sets of NNPDF~3.0~\cite{Butterworth:2015oua}.
\item \textit{\PQb tagging}: The uncertainties in the \PQb tagging and mistagging efficiency measurements are split into different components and propagated to the efficiency of tagging \PQb jets.
\end{itemize}
Nonprofiled uncertainties
\begin{itemize}
\item \textit{Luminosity}: The integrated luminosity is known with a relative uncertainty of {\tolerance=800 $\pm 2.6 \%$~\cite{CMS-PAS-LUM-17-001}.\par}
\item \textit{Jet energy scale (JES)}: All reconstructed jet four-momenta in simulated events are simultaneously varied according to the $\eta$- and $\pt$-dependent uncertainties in the JES~\cite{Khachatryan:2016kdb}. This variation in jet four-momenta is also propagated to $\ptmiss$.
\item \textit{Jet energy resolution (JER)}: A smearing is applied to account for the difference in the JER between simulation and data~\cite{Khachatryan:2016kdb}, and its uncertainty is estimated by increasing or decreasing the resolutions by their uncertainties.
\item \textit{Signal modelling}: The following uncertainty sources cover potential mismodelling of the single top quark $t$-channel signal processes. The effect of those uncertainties on the acceptance, and not on the cross section, is considered.
In the fit procedure, the uncertainties are not considered as nuisance parameters in the fit but evaluated by repeating the full analysis using samples of simulated signal events that feature variations in the modelling parameters covering the systematic uncertainty sources under study. 
\begin{itemize}
\item \textit{Signal $\mu_{\mathrm{R}}/\mu_{\mathrm{F}}$:} The uncertainties caused by variations in the renormalisation and factorisation scales are considered by reweighting the BDT response distributions according to weights corresponding to doubling/halving the nominal values of the scales~\cite{fourfiveflavorschemes,Alioli:2009je}.
\item \textit{Matching of matrix element and parton shower (ME-PS matching):} The parameter that controls the matching between the matrix element level calculation and the parton shower, and that regulates the high-\pt radiation in the simulation is varied within its uncertainties.
\item \textit{Parton shower factorisation scale:} The renormalisation scales of the initial- and final-state parton shower are varied by factors of two and one half with respect to the nominal value of 172.5\GeV.
\item \textit{PDF for signal process}: The uncertainty due to the choice of PDF is estimated using reweighted histograms derived from all PDF sets of NNPDF~3.0. The measurements in the following report only the experimental uncertainties, while the uncertainties on the predicted cross sections are reported in Table~\ref{tab:abgfit}. Effects on the fit due to correlation between PDFs are considered negligible.
\end{itemize}
\end{itemize}

\section{Fit procedure}
\label{sec:fit}
\label{sec:ckmfit}
The three CKM matrix elements are extracted by measuring the production cross sections and branching fractions of single top quark $t$-channel processes that depend on $\vtb$, $\vtd$, and $\vts$ in production and decay. The vast majority of single top quark $t$-channel events come from the $\STbb$ process, while $\STbq$ and $\STqb$ constitute subdominant production mechanisms. The $\TTqb$ contribution is taken into account in the background estimation.

The fit procedure is divided into two steps. 
In the first step, a maximum likelihood (ML) fit to the \mtw distribution is performed separately for the \twoJoneT and \threeJoneT categories in order to extract the QCD multijet contribution. The QCD multijet normalisation and the relative uncertainty are extrapolated to the QCD multijet depleted categories and used as an input to the second step.
In the second step, in order to discriminate between $\STbb$, $\STqb$, and $\STbq$, the multivariate discriminators described in Section~\ref{sec:ckmdiscr} are used in a simultaneous ML fit to the three event categories, while the QCD multijet prior uncertainty and central value are taken from the first step.

\begin{figure*}[!p]
\centering
\includegraphics[width=0.45\textwidth]{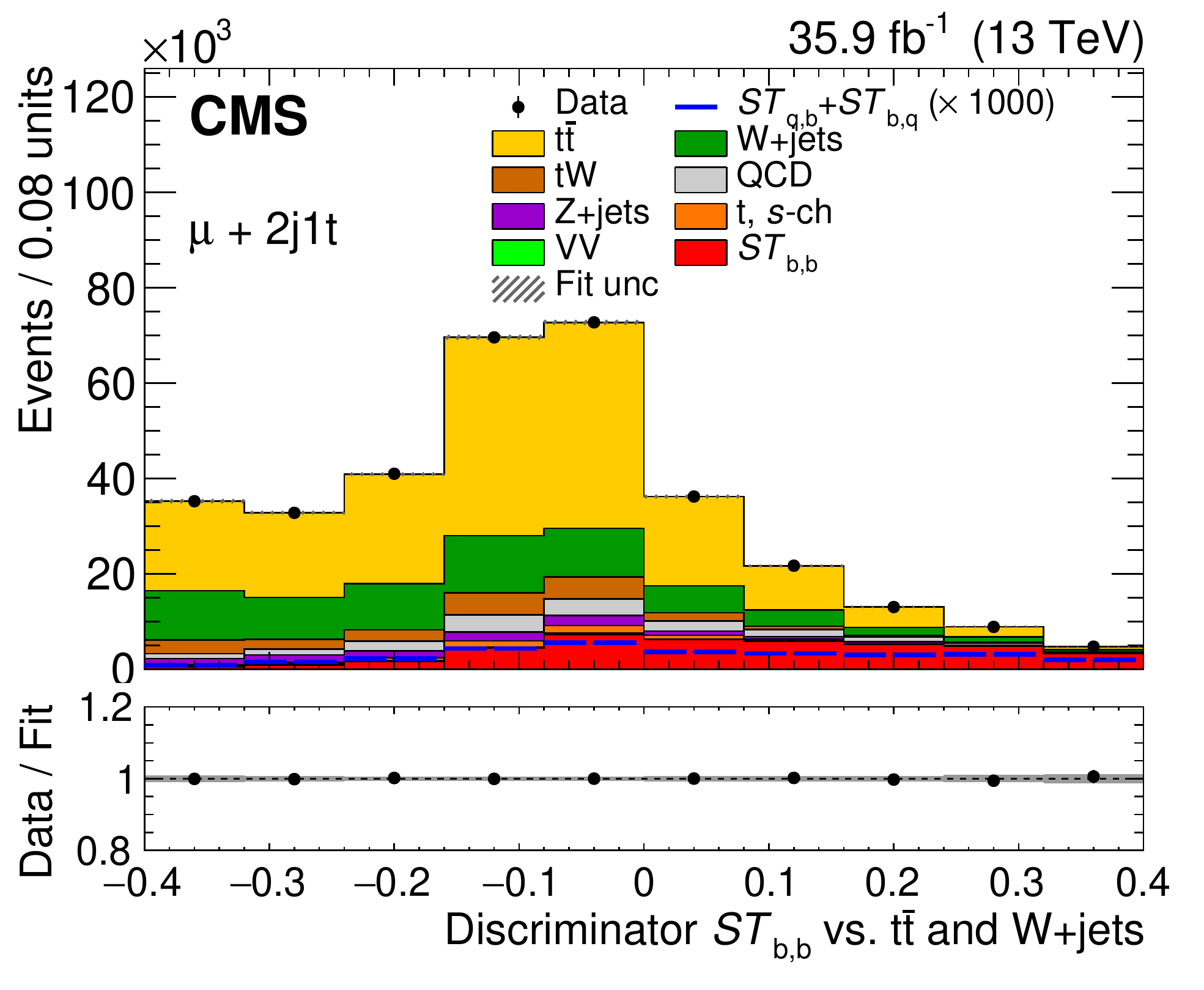}
\hspace{0.05\textwidth}
\includegraphics[width=0.45\textwidth]{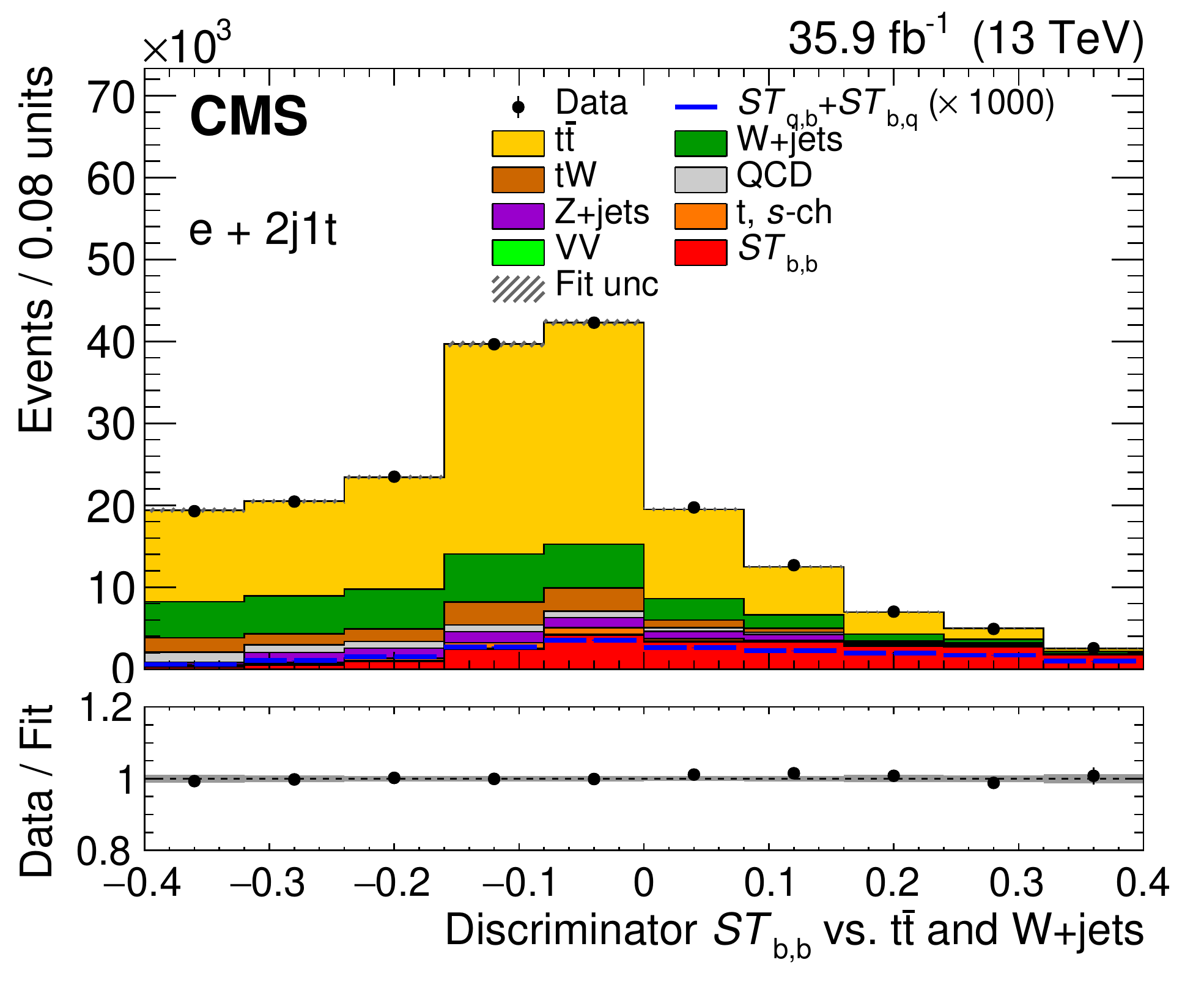}\\
\includegraphics[width=0.45\textwidth]{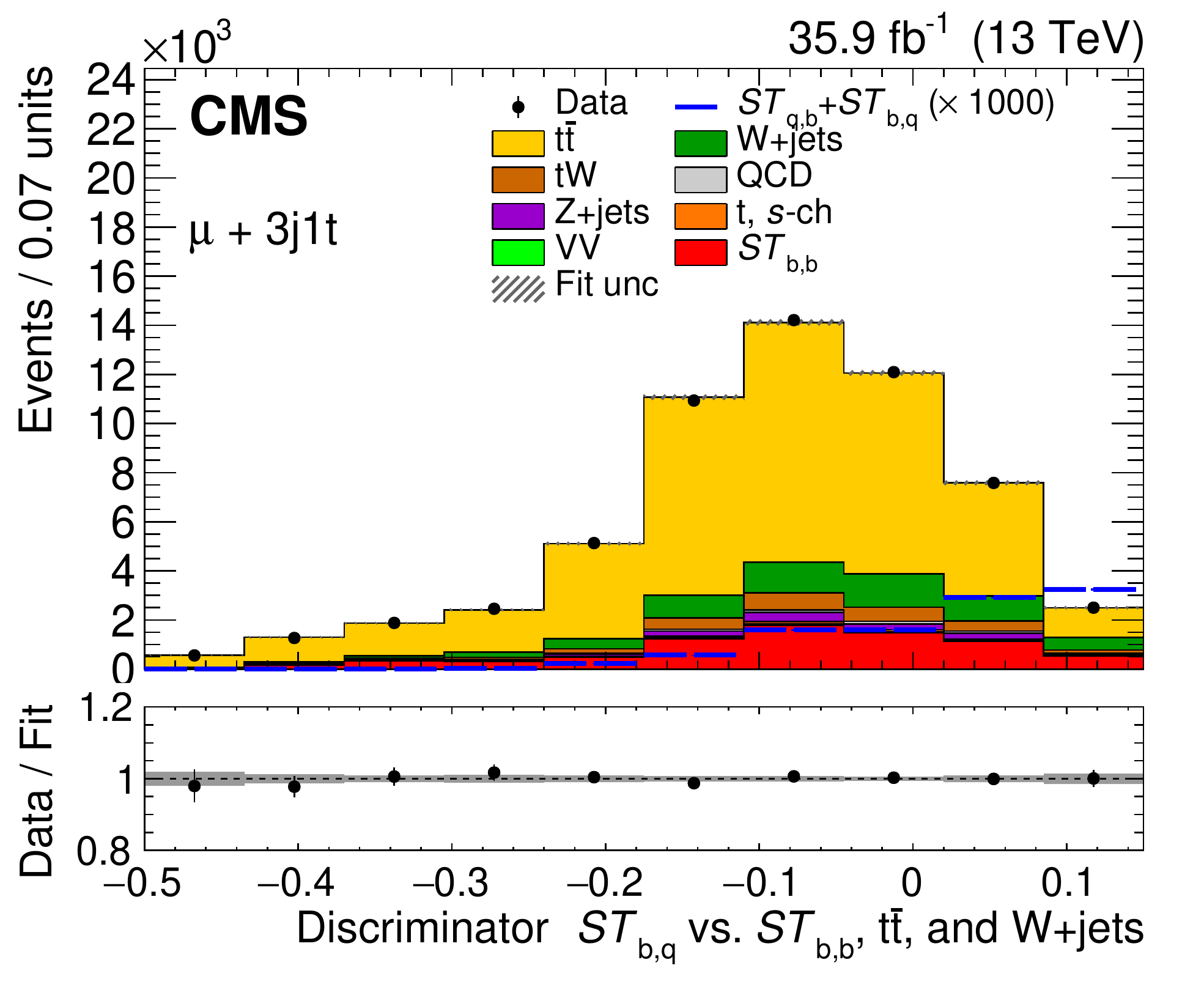}
\hspace{0.05\textwidth}
\includegraphics[width=0.45\textwidth]{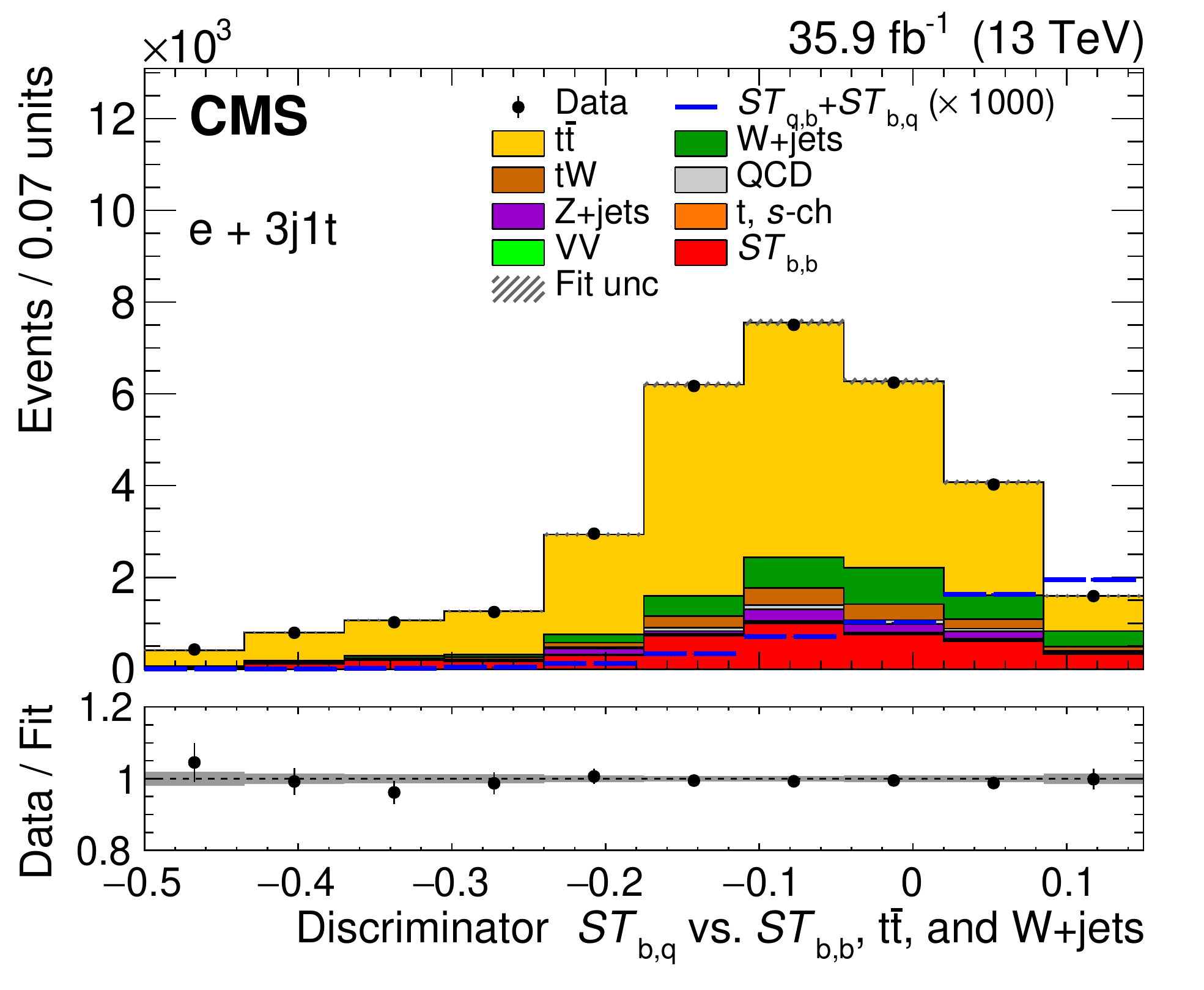}\\
\includegraphics[width=0.45\textwidth]{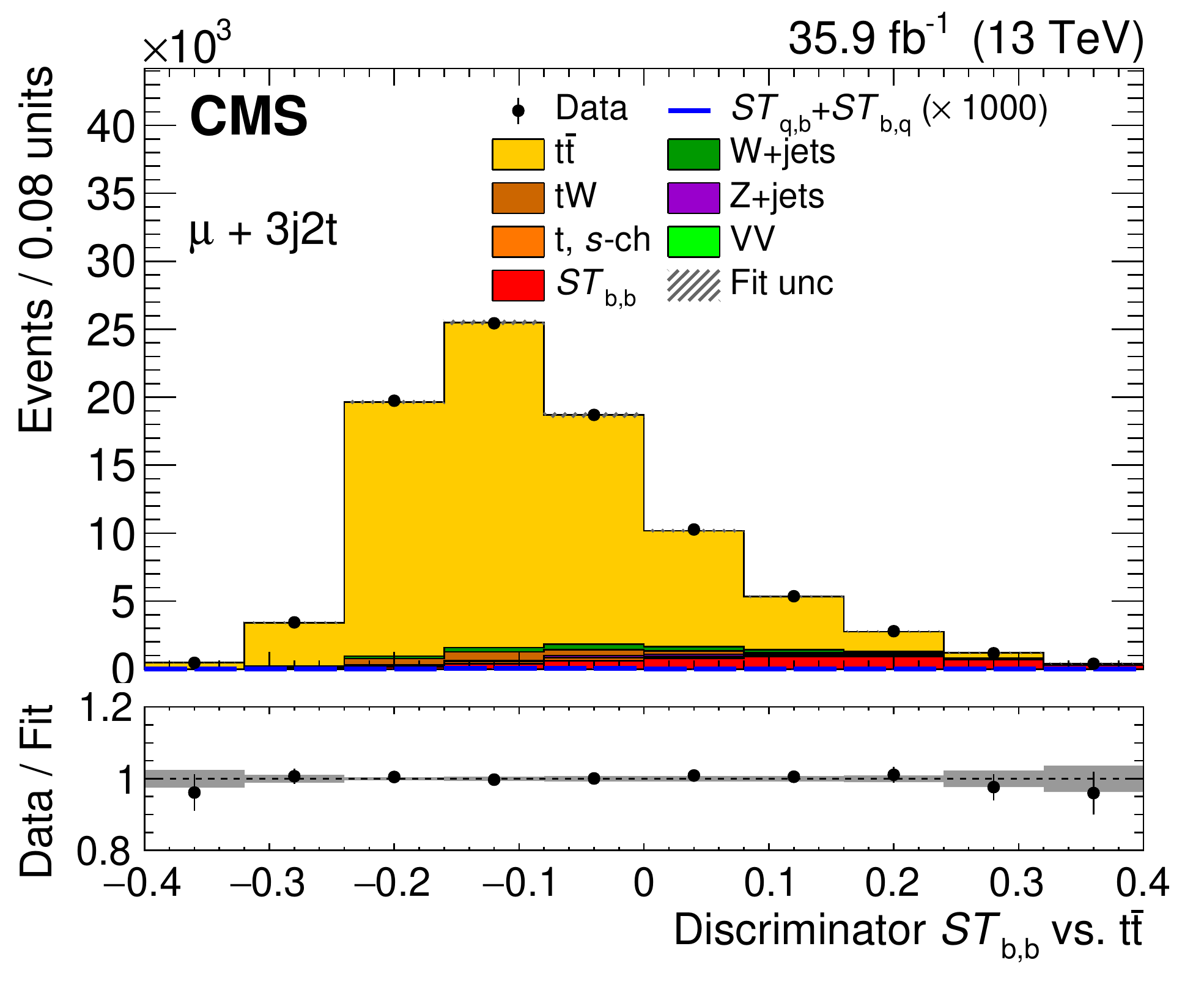}
\hspace{0.05\textwidth}
\includegraphics[width=0.45\textwidth]{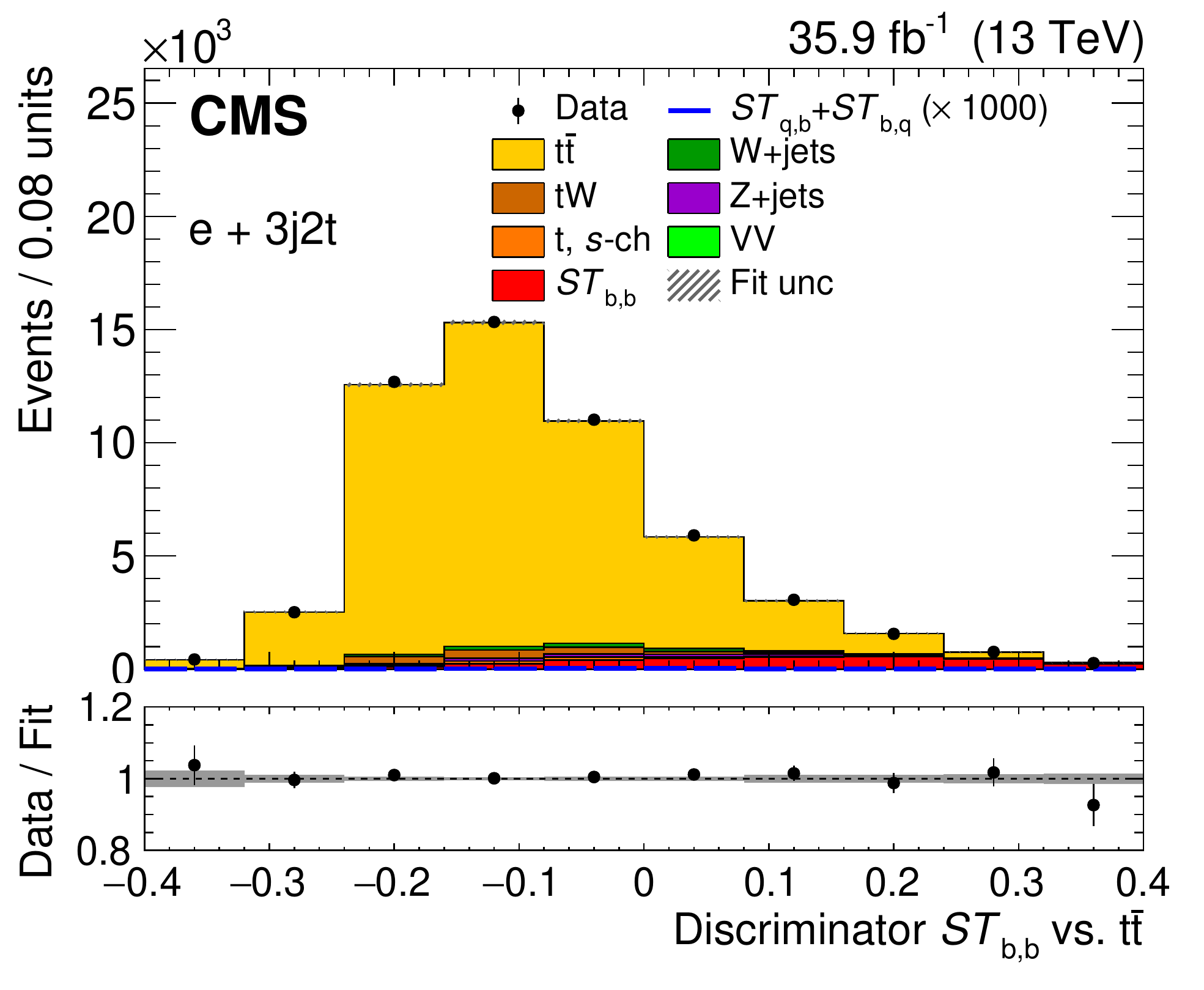}\\
\caption{\label{fig:stacks_postfit} Distribution of the multivariate discriminators, comparing data to simulation normalised after the fit procedure, for the muon channel on the left and for the electron channel on the right, for \twoJoneT(upper), \threeJoneT(middle), and \threeJtwoT(lower). The vertical lines on the points and the hatched bands show the experimental and fit uncertainties, respectively. The expected distribution from the $\STqb + \STbq$ processes (multiplied by a factor of 1000) is shown by the solid blue line. The lower panels show the ratio of the data to the fit.}
\end{figure*}

The $t$-channel single top quark signals are parametrised with a flat prior representing the coupling strength, and all systematic uncertainties  are treated as described in Section~\ref{sec:systs}. The smaller background yields are allowed to vary in the fit, along with the respective scale uncertainties. The QCD multijet background is fitted with a flat prior nuisance, while $\ttbar$ and \wjets backgrounds are left floating within the respective systematic uncertainties. 
The $t$-channel $\STbq$ and $\TTqb$ processes do not distinguish between topologies depending on  $\vtd$ or $\vts$ in the decay, while $\STqb$ is sensitive to the different PDFs contributing to the processes.
Figure \ref{fig:stacks_postfit} shows the distributions after the fit procedure has been applied for the muon (left) and the electron (right) channels.
The partial and total contributions of the profiled and nonprofiled uncertainties are given in Table~\ref{tab:systematics}.
\begin{table*}[!ht]
\centering
\topcaption{\label{tab:systematics} The sources and relative values in percent of the systematic uncertainty in the measurement of the \STbb cross section.  The uncertainties are broken up into profiled and nonprofiled sources.}
\begin{tabular}{llc}
Treatment & Uncertainty & $\Delta \sigma_{\STbb} / \sigma$ (\%) \\ \hline 
\multirow{10}{*}{Profiled}  & Lepton trigger and reconstruction & 0.50 \\
& Limited size of simulated event samples & 3.13 \\
& \ttbar modelling & 0.66 \\
& Pileup & 0.35 \\
& QCD background normalisation & 0.08 \\
& \wjets composition & 0.13 \\
& Other backgrounds $\mu_{\mathrm{R}}/\mu_{\mathrm{F}}$ & 0.44 \\
& PDF for background processes & 0.42 \\
& \PQb tagging & 0.73 \\ [\cmsTabSkip]
& Total profiled & 3.4 \\ [\cmsTabSkip]
\multirow{8}{*}{Nonprofiled} & Integrated luminosity & 2.5\\
& JER & 2.8\\
& JES & 8.0\\
& PDF for signal process & 3.8\\
& Signal $\mu_{\mathrm{R}}/\mu_{\mathrm{F}}$ &  2.4 \\
& ME-PS matching & 3.7\\
& Parton shower scale & 6.1\\  [\cmsTabSkip]
& Total nonprofiled & 11.5 \\ [\cmsTabSkip]
\multicolumn{2}{l}{Total uncertainty} & 12.0 \\
\end{tabular}
\end{table*}

\section{Results and interpretation}
\label{sec:results}

The contributions of each of the three CKM matrix elements to the different $\STbb$, $\STbq$, and $\STqb$ cross sections, extracted from the fit procedure, are considered.
In the SM, top quarks only decay to \PW bosons plus \PQb, \PQs, or \PQd quarks, and their branching fractions are proportional to the magnitude squared of the respective matrix element, as given in Table~\ref{tab:abgfit}. 
The fit results are given in terms of two signal strength parameters: the first, $\mu_{\PQb}$, refers to the $\STbb$ process, and the second, $\mu_{\PQs\PQd}$, to the sum of the $\STqb$ and $\STbq$ contributions. 

By neglecting terms proportional to $\abs{V_{\PQt\PQd}}^{4}$, $\abs{V_{\PQt\PQs}}^{4}$,  the $\STqb$ term can be written as proportional to $\absvtdsq+\absvtssq$, with a contribution of order 5\% that depends on $\absvtdsq/\absvtssq$. We consider variations on the latter contribution as negligible in the analysis. These assumptions can be justified because of the hierarchy observed in the first two rows of the CKM matrix. The signal strengths thus become:
\begin{linenomath}
\begin{equation}
\label{eq:masterrelations}
\begin{aligned}
\mu_{\PQb} & = \frac{\sigmatb^{\text{obs}} \BRtb^{\text{obs}}}{\sigmatb \BRtb} \\
\mu_{\PQs\PQd} & = \frac{\sigmatb^{\text{obs}} \BRtsd^{\text{obs}} + \sigmatsd^{\text{obs}}\BRtb^{\text{obs}}}{\sigmatb \BRtsd + \sigmatsd \BRtb},
\end{aligned}
\end{equation}
\end{linenomath}
where $\BRtsd$ is the branching fraction for a top quark to decay to a \PW boson and either an \PQs or \PQd quark. Henceforth, the ``obs'' label will refer to the measured value of a quantity, and the absence of this label will mean the expected value. Equation~(\ref{eq:masterrelations}) shows that the signal strengths are the ratios of the measured value of a quantity to the expected value.

One can write Eq.~(\ref{eq:masterrelations}) more generally in terms of the top quark decay amplitudes or partial widths. We factorise out the modulus of the matrix element from the partial width for each quark. Thus, the top quark partial width to $\PW\PQq$ can be written as $\Gamma_{\PQq}=\widetilde{\Gamma}_{\PQq}\absvtqsq$, where $\widetilde{\Gamma}_{\PQq}$ is the top quark partial width for $\absvtq=1$. We further assume that $\widetilde{\Gamma}_{\PQq}=\widetilde{\Gamma}_{\PQb}$, \ie that any differences other than the CKM elements are negligible.
Using this and the total width $\Gamma_{\PQt}$ of the top quark, we can write Eq.~(\ref{eq:masterrelations}) as:
\begin{linenomath}
\begin{equation}
\label{master2}
\begin{aligned}
\mu_{\PQb} & = \frac{\absvtbfourth_{\text{obs}} \widetilde{\Gamma}_{\PQq}^{\text{obs}} \Gamma_{\PQt}} {\absvtbfourth \widetilde{\Gamma}_{\PQq} \Gamma_{\PQt}^{\text{obs} } } \\
\mu_{\PQs\PQd} & = \frac{\absvtbsq_{\text{obs}} \bigl(\absvtssq_{\text{obs}} + \absvtdsq_{\text{obs}}\bigr) \widetilde{\Gamma}_{\PQq}^{\text{obs}} \Gamma_{\PQt}}{\absvtbsq \bigl(\absvtssq + \absvtdsq\bigr) \widetilde{\Gamma}_{\PQq} \Gamma_{\PQt}^{\text{obs}}}.
\end{aligned}
\end{equation}
\end{linenomath}

The first fit extracts the signal strengths $\mu_{\PQb}$ and $\mu_{\PQs\PQd}$, whose values can be interpreted under different model assumptions. The signal strengths obtained are:
\begin{linenomath}
\begin{equation}
\begin{aligned}
\mu_{\PQb} & = 0.99 \pm 0.03\,  (\text{stat+prof}) \pm 0.12\,  (\text{nonprof}) \\
\mu_{\PQs\PQd} &  < 87\ \text{at 95\% confidence level (\CL),}
\end{aligned}
\end{equation}
\end{linenomath}
with a correlation factor of $\rho_{\mu_{\PQb}, \mu_{\PQs\PQd}} = -0.25$. The first uncertainty on $\mu_{\PQb}$ is the combination of the statistical and profiled systematic uncertainties, while the second is due to the nonprofiled systematic components. The upper limit on $\mu_{\PQs\PQd}$ takes into account both profiled and nonprofiled systematic uncertainties.

In the following, we describe the signal extraction using the values of the CKM elements directly as parameters in the fit and applying constraints from the SM scenario and then two possible beyond-the-SM (BSM) extensions.

\subsection{Measurement in the SM scenario}
One can simplify Eq.~(\ref{eq:masterrelations}) by assuming the SM unitarity constraint $\absvtbsq +\absvtdsq + \absvtssq=1$. The fit is repeated, taking \absvtb as the single free parameter and replacing $\absvtdsq +\absvtssq$ with $1- \absvtbsq$. 
In this case, Eq.~(\ref{eq:masterrelations}) becomes:
\begin{linenomath}
\begin{equation}
\label{smconstrainedmaster}
\begin{aligned}
\mu_{\PQb} & = \frac{\absvtbfourth_{\text{obs} }}{\absvtbfourth } \\
\mu_{\PQs\PQd} & = \frac{\absvtbsq_{\text{obs}} \bigl(1- \absvtbsq_{\text{obs}} \bigr) }{\absvtbsq \bigl(1- \absvtbsq\bigr)}.
\end{aligned}
\end{equation}
\end{linenomath}

The fit is only allowed to return values of $\absvtb\,\le\,1$, and the constraint $\absvtdsq+\absvtssq=1-\absvtbsq$ is imposed. Because of these constraints, Gaussian behaviour of the uncertainties cannot be assumed. Instead, pseudo-experiments are generated to evaluate the impact of nonprofiled uncertainties on the measurement, and the following confidence intervals are measured at 95\% \CL:
\begin{linenomath}
\begin{equation}
\label{smconstrained}
\begin{aligned}
& \absvtb \, > 0.970 \\ 
& \absvtdsq +\absvtssq < 0.057.\\
\end{aligned}
\end{equation}
\end{linenomath}
This measurement is comparable with the previous most precise estimate using \ttbar events from Ref.~\cite{Khachatryan:2014nda}, and with the result of the combination of single top quark measurements in Ref.~\cite{Aaboud:2019pkc}.

\subsection{Measurements for two BSM scenarios}
Any BSM contribution potentially enhancing $\absvtbsq$, $\absvtssq$, or $\absvtdsq$ can affect top quark production, decay, or both. 
Some BSM scenarios predict the presence of additional quark families. In this case, the CKM matrix is extended due to the mixing between the SM quarks and the new hypothesised ones. This would imply that the CKM matrix elements $\absvtb$, $\absvts$, and $\absvtd$ would not necessarily satisfy the unitarity constraint of $\absvtbsq + \absvtssq + \absvtdsq=1$.
If these BSM quarks are heavier than the top quark, they would alter the CKM matrix elements without appearing as top quark decay products. They would thus not contribute directly to the top quark decay width $\Gamma_{\PQt}$, but only indirectly because of the reduction in the absolute values of the corresponding SM CKM matrix elements.

For the first BSM scenario, we assume the top quark decays through the same channels as in the SM case, and that the partial width of each decay only varies because of a modified CKM matrix element. In this case, by writing $\Gamma_{\PQt}$ and $\widetilde{\Gamma}_{\PQq}$  as a function of $\absvtbsq$ and $\absvtdsq+\absvtssq$, Eq.~(\ref{master2}) becomes:
\begin{linenomath}
\begin{equation}
\label{bsmconstrainedmaster}
\begin{aligned}
\mu_{\PQb} & = \frac{\absvtbfourth_{\text{obs}}}{\absvtbfourth \bigl(\absvtbsq_{\text{obs}} + \absvtssq_{\text{obs}} + \absvtdsq_{\text{obs}}\bigr)} \\
\mu_{\PQs\PQd} & = \frac{\absvtbsq_{\text{obs}} \bigl(\absvtssq_{\text{obs}} + \absvtdsq_{\text{obs}}\bigr)}{\bigl(\absvtssq + \absvtdsq\bigr) \bigl(\absvtbsq_{\text{obs}} + \absvtssq_{\text{obs}} + \absvtdsq_{\text{obs}}\bigr)}. 
\end{aligned}
\end{equation}
\end{linenomath}

In this scenario, the measurement is performed leaving \absvtb and \absvtdsq +\absvtssq as free parameters in the fit, resulting in:
\begin{linenomath}
\begin{equation}
\label{bsmconstrained}
\begin{aligned}
& \absvtb = 0.988 \pm 0.027\, (\text{stat+prof})  \pm 0.043 \,(\text{nonprof} )\\ 
& \absvtdsq +\absvtssq = 0.06 \pm 0.05 \, (\text{stat+prof})\, ^{+0.04}_{-0.03}\, (\text{nonprof}).
\end{aligned}
\end{equation}
\end{linenomath}

In the second BSM scenario, the top quark partial width is unchanged, but the total width increases due to additional, undetected decays. In the fit, the partial widths for decays to known quarks are fixed, and the total width
is a free parameter and allowed to vary. The effects on $\Gamma_{\PQt}$ due to variations in \absvtbsq, \absvtdsq, and \absvtssq are neglected.

In this scenario, Eq.~(\ref{master2}) is modified to:
\begin{linenomath}
\begin{equation}
\label{bsmgeneric}
\begin{aligned}
\mu_{\PQb} & = \frac{\absvtbfourth_{\text{obs}} \Gamma_{\PQt}}{\absvtbfourth \Gamma_{\PQt}^{\text{obs}}}\\
\mu_{\PQs\PQd} & = \frac{\absvtbsq_{\text{obs} } \bigl(\absvtssq_{\text{obs}} + \absvtdsq_{\text{obs} } \bigr) \Gamma_{\PQt} }{ \absvtbsq \bigl( \absvtssq + \absvtdsq \bigr) \Gamma_{\PQt}^{\text{obs} } }.
\end{aligned}
\end{equation}
\end{linenomath}

Using $\absvtbsq$, $\absvtdsq + \absvtssq$, and  $R_{\Gamma}=\Gamma_{\PQt}^{\text{obs}}/\Gamma_{\PQt}$ as the free parameters in the fit, we obtain:
\begin{linenomath}
\begin{equation}
\label{bsmconstrainedgamma}
\begin{aligned}
& \absvtb = 0.988 \pm 0.011 \, (\text{stat+prof})  \pm 0.021 \, (\text{nonprof} ) \\ 
& \absvtdsq +\absvtssq = 0.06 \pm 0.05 \, (\text{stat+prof}) \pm 0.04 \, (\text{nonprof}) \\
& R_{\Gamma} = 0.99 \pm 0.42 \, (\text{stat+prof}) \pm 0.03 \, (\text{nonprof}).
\end{aligned}
\end{equation}
\end{linenomath}
The measured correlation factors between the three parameters are  $\rho_{\absvtb,\absvtdsq} = -0.19$, $\rho_{\absvtb,R_{\Gamma}} = -0.78$, and $\rho_{R_{\Gamma},\absvtdsq} = -0.21$. This measurement is in good agreement with the other measurements from Refs.~\cite{Abazov:2012vd,Khachatryan:2014nda,Aaboud:2017uqq,Aaboud:2019pkc}, which however make use of the SM assumptions.
The results for the second BSM scenario have a higher statistical precision than those for the first scenario because of the weaker dependence of the signal strength on \absvtb for the first scenario.

As mentioned in Section \ref{sec:introduction}, constraints on \absvtd and \absvts from precision low-energy measurements do not necessarily hold when BSM particles are present in the relevant Feynman diagram loops. Theoretical studies have shown that values of \absvts up to about $0.2$ are possible in some BSM scenarios~\cite{Alwall:2006bx}. The measurements presented here establish a model-independent upper limit on \absvtd and \absvts by removing any assumed theoretical hypotheses. This will now allow new interpretations for possible mixing of SM and BSM processes.

Alternative approaches interpret the available single top quark measurements in terms of different scenarios for modifying the CKM matrix elements (see, for example, Ref.~\cite{Clerbaux:2018vup}), obtaining results that are comparable with the measurements presented in this Letter. Such approaches, however, do not allow changes in the decay vertex of the top quark, and do not consider possible similarities in the features of the~$\STbq$ signal and background processes.

The current analysis improves the precision on $\absvtb$ by 50\% with respect to previous studies~\cite{Sirunyan:2018rlu} by exploiting the $\PQt\PW\PQb$ vertex in the top quark decay, and is more precise than the combined ATLAS and CMS measurement using data at $\sqrt{s} = 7 $ and 8\TeV~\cite{Aaboud:2019pkc}.

\section{Summary}
\label{sec:summary}
A measurement of the Cabibbo--Kobayashi--Maskawa (CKM) matrix elements $\absvtb$, $\absvtd$, and $\absvts$ has been performed in an event sample enriched in $t$-channel single top quark events, featuring one muon or electron and jets in the final state. The data are from proton-proton collisions at $\sqrt{s} = 13\TeV$, acquired at the LHC by the CMS experiment and correspond to an integrated luminosity of $35.9\fbinv$.
The contributions from single top quark processes featuring all three matrix elements in the production vertex have been considered as separate signal processes, as well as contributions from decays of single top quarks involving all three quark families. 
The yields of the signal processes have been extracted through a simultaneous fit to data in different selected event categories, and the values of the CKM matrix elements have been inferred from the signal strengths, which are the ratios of the measured top quark $t$-channel cross sections times branching ratios to the expected values.
The signal strengths obtained from the fit are $\mu_{\PQb} = 0.99 \pm 0.12$, where the uncertainty includes both the statistical and systematic components, and $\mu_{\PQs\PQd} < 87\ \mathrm{at}\ 95\%$ confidence level (\CL).

Under the standard model assumption of CKM unitarity, the values are found to be $ \absvtb > 0.970$ and $ \absvtdsq +\absvtssq < 0.057$, both at $95\%$ \CL.

Fits were also performed under two different beyond-the-standard-model scenarios.  In the first, we assume the presence of additional quark families that are heavier than the top quark. The unitarity constraint for the three CKM matrix elements no longer holds, but the top quark decays through the same channels as in the standard model.  We assume the partial width of each top quark decay only varies because of a modified CKM matrix element. The fit gives:
\begin{linenomath}
\begin{equation*}
\begin{aligned}
& \absvtb = 0.988 \pm 0.051\\
& \absvtdsq +\absvtssq = 0.06 \pm 0.06,\\
\end{aligned}
\end{equation*}
\end{linenomath}
where the uncertainties include both the statistical and systematic components.

In the second scenario, the top quark width is left unconstrained under the assumption that the contributions to the total width from the mixing of the three families are negligible. The corresponding measured values are:
\begin{linenomath}
\begin{equation*}
\begin{aligned}
& \absvtb = 0.988 \pm 0.024,\\
& \absvtdsq +\absvtssq = 0.06 \pm 0.06,\\
& \frac{\Gamma^{\text{obs}}_{\mathrm{t}}}{\Gamma_{\PQt}} = 0.99 \pm 0.42,
\end{aligned}
\end{equation*}
\end{linenomath}
where again, both the statistical and systematic uncertainties are included. The differences among the uncertainties in the presented scenarios are driven by the difference in the functional dependence of the observed event yields from the CKM matrix elements. This results in smaller uncertainties in \absvtb for the case where a fourth--power dependence is considered with respect to the second--power dependence case.

All results are consistent with each other, and show no deviation with respect to extrapolations of low-energy measurements. These results are the first direct, model-independent measurements of the CKM matrix elements for the third-generation quarks, and provide the best determination of these fundamental SM parameters via single top quark measurements.

\begin{acknowledgments}

  We congratulate our colleagues in the CERN accelerator departments for the excellent performance of the LHC and thank the technical and administrative staffs at CERN and at other CMS institutes for their contributions to the success of the CMS effort. In addition, we gratefully acknowledge the computing centres and personnel of the Worldwide LHC Computing Grid for delivering so effectively the computing infrastructure essential to our analyses. Finally, we acknowledge the enduring support for the construction and operation of the LHC and the CMS detector provided by the following funding agencies: BMBWF and FWF (Austria); FNRS and FWO (Belgium); CNPq, CAPES, FAPERJ, FAPERGS, and FAPESP (Brazil); MES (Bulgaria); CERN; CAS, MoST, and NSFC (China); COLCIENCIAS (Colombia); MSES and CSF (Croatia); RPF (Cyprus); SENESCYT (Ecuador); MoER, ERC IUT, PUT and ERDF (Estonia); Academy of Finland, MEC, and HIP (Finland); CEA and CNRS/IN2P3 (France); BMBF, DFG, and HGF (Germany); GSRT (Greece); NKFIA (Hungary); DAE and DST (India); IPM (Iran); SFI (Ireland); INFN (Italy); MSIP and NRF (Republic of Korea); MES (Latvia); LAS (Lithuania); MOE and UM (Malaysia); BUAP, CINVESTAV, CONACYT, LNS, SEP, and UASLP-FAI (Mexico); MOS (Montenegro); MBIE (New Zealand); PAEC (Pakistan); MSHE and NSC (Poland); FCT (Portugal); JINR (Dubna); MON, RosAtom, RAS, RFBR, and NRC KI (Russia); MESTD (Serbia); SEIDI, CPAN, PCTI, and FEDER (Spain); MOSTR (Sri Lanka); Swiss Funding Agencies (Switzerland); MST (Taipei); ThEPCenter, IPST, STAR, and NSTDA (Thailand); TUBITAK and TAEK (Turkey); NASU (Ukraine); STFC (United Kingdom); DOE and NSF (USA). 
 
  \hyphenation{Rachada-pisek} Individuals have received support from the Marie-Curie programme and the European Research Council and Horizon 2020 Grant, contract Nos.\ 675440, 752730, and 765710 (European Union); the Leventis Foundation; the A.P.\ Sloan Foundation; the Alexander von Humboldt Foundation; the Belgian Federal Science Policy Office; the Fonds pour la Formation \`a la Recherche dans l'Industrie et dans l'Agriculture (FRIA-Belgium); the Agentschap voor Innovatie door Wetenschap en Technologie (IWT-Belgium); the F.R.S.-FNRS and FWO (Belgium) under the ``Excellence of Science -- EOS" -- be.h project n.\ 30820817; the Beijing Municipal Science \& Technology Commission, No. Z191100007219010; the Ministry of Education, Youth and Sports (MEYS) of the Czech Republic; the Deutsche Forschungsgemeinschaft (DFG) under Germany's Excellence Strategy -- EXC 2121 ``Quantum Universe" -- 390833306; the Lend\"ulet (``Momentum") Programme and the J\'anos Bolyai Research Scholarship of the Hungarian Academy of Sciences, the New National Excellence Program \'UNKP, the NKFIA research grants 123842, 123959, 124845, 124850, 125105, 128713, 128786, and 129058 (Hungary); the Council of Science and Industrial Research, India; the HOMING PLUS programme of the Foundation for Polish Science, cofinanced from European Union, Regional Development Fund, the Mobility Plus programme of the Ministry of Science and Higher Education, the National Science Center (Poland), contracts Harmonia 2014/14/M/ST2/00428, Opus 2014/13/B/ST2/02543, 2014/15/B/ST2/03998, and 2015/19/B/ST2/02861, Sonata-bis 2012/07/E/ST2/01406; the National Priorities Research Program by Qatar National Research Fund; the Ministry of Science and Education, grant no. 14.W03.31.0026 (Russia); the Tomsk Polytechnic University Competitiveness Enhancement Program and ``Nauka" Project FSWW-2020-0008 (Russia); the Programa Estatal de Fomento de la Investigaci{\'o}n Cient{\'i}fica y T{\'e}cnica de Excelencia Mar\'{\i}a de Maeztu, grant MDM-2015-0509 and the Programa Severo Ochoa del Principado de Asturias; the Thalis and Aristeia programmes cofinanced by EU-ESF and the Greek NSRF; the Rachadapisek Sompot Fund for Postdoctoral Fellowship, Chulalongkorn University and the Chulalongkorn Academic into Its 2nd Century Project Advancement Project (Thailand); the Kavli Foundation; the Nvidia Corporation; the SuperMicro Corporation; the Welch Foundation, contract C-1845; and the Weston Havens Foundation (USA). 
\end{acknowledgments}

\bibliography{auto_generated} 

\cleardoublepage \appendix\section{The CMS Collaboration \label{app:collab}}\begin{sloppypar}\hyphenpenalty=5000\widowpenalty=500\clubpenalty=5000\input{TOP-17-012-authorlist.tex}\end{sloppypar}
%%% END EDITABLE REGION %%%
\end{document}

%% file: TOP-17-012-authorlist.tex
\vskip\cmsinstskip
\textbf{Yerevan Physics Institute, Yerevan, Armenia}\\*[0pt]
A.M.~Sirunyan$^{\textrm{\dag}}$, A.~Tumasyan
\vskip\cmsinstskip
\textbf{Institut f\"{u}r Hochenergiephysik, Wien, Austria}\\*[0pt]
W.~Adam, F.~Ambrogi, T.~Bergauer, M.~Dragicevic, J.~Er\"{o}, A.~Escalante~Del~Valle, M.~Flechl, R.~Fr\"{u}hwirth\cmsAuthorMark{1}, M.~Jeitler\cmsAuthorMark{1}, N.~Krammer, I.~Kr\"{a}tschmer, D.~Liko, T.~Madlener, I.~Mikulec, N.~Rad, J.~Schieck\cmsAuthorMark{1}, R.~Sch\"{o}fbeck, M.~Spanring, W.~Waltenberger, C.-E.~Wulz\cmsAuthorMark{1}, M.~Zarucki
% ...
\vskip\cmsinstskip
\dag: Deceased\\
1:  Also at Vienna University of Technology, Vienna, Austria\\
% ...


%% file: TOP-17-012_temp.bbl
\providecommand{\href}[2]{#2}\begingroup\raggedright\begin{thebibliography}{10}%
\makeatletter
\providecommand{\hrefCMSnoop }[0]{\@secondoftwo}%
\makeatother
\providecommand{\doi}{\texttt{doi:}\begingroup \urlstyle{tt}\Url}

\bibitem{Kobayashi:1973fv}
\hrefCMSnoop {}{M.~Kobayashi and T.~Maskawa, ``{CP-violation in the
  renormalizable theory of weak interaction}'',} \textit{ Prog. Theor. Phys.}
  \textbf{ 49} (1973) 652,
\href{http://dx.doi.org/10.1143/PTP.49.652}{\doi{10.1143/PTP.49.652}}.
%%CITATION = PTPKA,49,652;%%.

\bibitem{Chatrchyan:2011vp}
\hrefCMSnoop {}{{CMS Collaboration}, ``{Measurement of the $t$-channel single
  top quark production cross section in $pp$ collisions at $\sqrt{s}=7$
  TeV}'',} \textit{ Phys. Rev. Lett.} \textbf{ 107} (2011) 091802,
  \href{http://dx.doi.org/10.1103/PhysRevLett.107.091802}{\doi{10.1103/PhysRevLett.107.091802}},
  \href{http://www.arXiv.org/abs/1106.3052}{\texttt{arXiv:1106.3052}}.

\bibitem{Chatrchyan:2012ep}
\hrefCMSnoop {}{{CMS Collaboration}, ``{Measurement of the single-top-quark
  $t$-channel cross section in pp collisions at $\sqrt{s}= $ 7 \TeV}'',}
  \textit{ JHEP} \textbf{ 12} (2012) 035,
  \href{http://dx.doi.org/10.1007/JHEP12(2012)035}{\doi{10.1007/JHEP12(2012)035}},
\href{http://www.arXiv.org/abs/1209.4533}{\texttt{arXiv:1209.4533}}.
%%CITATION = ARXIV:1209.4533;%%.

\bibitem{Aad:2012ux}
\hrefCMSnoop {}{{ATLAS Collaboration}, ``{Measurement of the $t$-channel single
  top-quark production cross section in pp collisions at $\sqrt{s}=7$ TeV with
  the ATLAS detector}'',} \textit{ Phys. Lett. B} \textbf{ 717} (2012) 330,
  \href{http://dx.doi.org/10.1016/j.physletb.2012.09.031}{\doi{10.1016/j.physletb.2012.09.031}},
\href{http://www.arXiv.org/abs/1205.3130}{\texttt{arXiv:1205.3130}}.
%%CITATION = ARXIV:1205.3130;%%.

\bibitem{Aad:2014fwa}
\hrefCMSnoop {}{{ATLAS Collaboration}, ``{Comprehensive measurements of
  $t$-channel single top-quark production cross sections at $\sqrt{s} = 7$ TeV
  with the ATLAS detector}'',} \textit{ Phys. Rev. D} \textbf{ 90} (2014)
  112006,
  \href{http://dx.doi.org/10.1103/PhysRevD.90.112006}{\doi{10.1103/PhysRevD.90.112006}},
  \href{http://www.arXiv.org/abs/1406.7844}{\texttt{arXiv:1406.7844}}.

\bibitem{Khachatryan:2014iya}
\hrefCMSnoop {}{{CMS Collaboration}, ``{Measurement of the $t$-channel
  single-top-quark production cross section and of the $| V_{\rm tb} |$ CKM
  matrix element in pp collisions at $\sqrt{s}= 8$ TeV}'',} \textit{ JHEP}
  \textbf{ 06} (2014) 090,
  \href{http://dx.doi.org/10.1007/JHEP06(2014)090}{\doi{10.1007/JHEP06(2014)090}},
\href{http://www.arXiv.org/abs/1403.7366}{\texttt{arXiv:1403.7366}}.
%%CITATION = ARXIV:1403.7366;%%.

\bibitem{Aaboud:2016ymp}
\hrefCMSnoop {}{{ATLAS Collaboration}, ``{Measurement of the inclusive
  cross-sections of single top-quark and top-antiquark $t$-channel production
  in pp collisions at $\sqrt{s} = 13$ TeV with the ATLAS detector}'',} \textit{
  JHEP} \textbf{ 04} (2017) 086,
  \href{http://dx.doi.org/10.1007/JHEP04(2017)086}{\doi{10.1007/JHEP04(2017)086}},
  \href{http://www.arXiv.org/abs/1609.03920}{\texttt{arXiv:1609.03920}}.

\bibitem{Sirunyan:2016cdg}
\hrefCMSnoop {}{{CMS Collaboration}, ``{Cross section measurement of
  $t$-channel single top quark production in pp collisions at $\sqrt{s} = 13$
  TeV}'',} \textit{ Phys. Lett. B} \textbf{ 772} (2017) 752,
  \href{http://dx.doi.org/10.1016/j.physletb.2017.07.047}{\doi{10.1016/j.physletb.2017.07.047}},
\href{http://www.arXiv.org/abs/1610.00678}{\texttt{arXiv:1610.00678}}.
%%CITATION = ARXIV:1610.00678;%%.

\bibitem{Aaboud:2017pdi}
\hrefCMSnoop {}{{ATLAS Collaboration}, ``{Fiducial, total and differential
  cross-section measurements of $t$-channel single top-quark production in pp
  collisions at 8 TeV using data collected by the ATLAS detector}'',} \textit{
  Eur. Phys. J. C} \textbf{ 77} (2017) 531,
  \href{http://dx.doi.org/10.1140/epjc/s10052-017-5061-9}{\doi{10.1140/epjc/s10052-017-5061-9}},
  \href{http://www.arXiv.org/abs/1702.02859}{\texttt{arXiv:1702.02859}}.

\bibitem{Sirunyan:2018rlu}
\hrefCMSnoop {}{{CMS Collaboration}, ``{Measurement of the single top quark and
  antiquark production cross sections in the $t$ channel and their ratio in
  proton-proton collisions at $\sqrt{s}=$ 13 TeV}'',} \textit{ Phys. Lett. B}
  \textbf{ 800} (2019) 135042,
  \href{http://dx.doi.org/10.1016/j.physletb.2019.135042}{\doi{10.1016/j.physletb.2019.135042}},
\href{http://www.arXiv.org/abs/1812.10514}{\texttt{arXiv:1812.10514}}.
%%CITATION = ARXIV:1812.10514;%%.

\bibitem{Sirunyan:2019hqb}
\hrefCMSnoop {}{{CMS Collaboration}, ``{Measurement of differential cross
  sections and charge ratios for t-channel single top quark production in
  proton--proton collisions at $\sqrt{s}=13\,\text {Te}\text {V}$}'',} \textit{
  Eur. Phys. J. C} \textbf{ 80} (2020), no.~5, 370,
  \href{http://dx.doi.org/10.1140/epjc/s10052-020-7858-1}{\doi{10.1140/epjc/s10052-020-7858-1}},
  \href{http://www.arXiv.org/abs/1907.08330}{\texttt{arXiv:1907.08330}}.

\bibitem{PDG2018}
\hrefCMSnoop {}{{Particle Data Group}, M.~Tanabashi {et~al.}, ``Review of
  particle physics'',} \textit{ Phys. Rev. D} \textbf{ 98} (2018) 030001,
  \href{http://dx.doi.org/10.1103/PhysRevD.98.030001}{\doi{10.1103/PhysRevD.98.030001}}.

\bibitem{Alwall:2006bx}
J.~Alwall\hrefCMSnoop {}{ {et~al.}, ``{Is $V_{\mathrm{tb}}\simeq 1$?}'',}
  \textit{ Eur. Phys. J. C} \textbf{ 49} (2007) 791,
  \href{http://dx.doi.org/10.1140/epjc/s10052-006-0137-y}{\doi{10.1140/epjc/s10052-006-0137-y}},
  \href{http://www.arXiv.org/abs/hep-ph/0607115}{\texttt{arXiv:hep-ph/0607115}}.

\bibitem{Abazov:2011zk}
\hrefCMSnoop {}{{D0} Collaboration, ``{Precision measurement of the ratio $\rm
  \mathcal{B}(t \to Wb)/\mathcal{B}(t \to Wq)$ and extraction of $V_{\rm
  tb}$}'',} \textit{ Phys. Rev. Lett.} \textbf{ 107} (2011) 121802,
  \href{http://dx.doi.org/10.1103/PhysRevLett.107.121802}{\doi{10.1103/PhysRevLett.107.121802}},
\href{http://www.arXiv.org/abs/1106.5436}{\texttt{arXiv:1106.5436}}.
%%CITATION = ARXIV:1106.5436;%%.

\bibitem{Aaltonen:2013luz}
\hrefCMSnoop {}{{CDF} Collaboration, ``{Measurement of $R = \mathcal{B}({\rm t
  \rightarrow Wb})/\mathcal{B}({\rm t \rightarrow Wq})$ in top--quark--pair
  decays using lepton+jets events and the full CDF Run II data set}'',}
  \textit{ Phys. Rev. D} \textbf{ 87} (2013) 111101,
  \href{http://dx.doi.org/10.1103/PhysRevD.87.111101}{\doi{10.1103/PhysRevD.87.111101}},
\href{http://www.arXiv.org/abs/1303.6142}{\texttt{arXiv:1303.6142}}.
%%CITATION = ARXIV:1303.6142;%%.

\bibitem{Aaltonen:2014yua}
\hrefCMSnoop {}{{CDF} Collaboration, ``{Measurement of $\mathcal{B}({\rm t \to
  Wb})/\mathcal{B}({\rm t \to Wq})$ in top-quark-pair decays using dilepton
  events and the full CDF Run II data set}'',} \textit{ Phys. Rev. Lett.}
  \textbf{ 112} (2014) 221801,
  \href{http://dx.doi.org/10.1103/PhysRevLett.112.221801}{\doi{10.1103/PhysRevLett.112.221801}},
\href{http://www.arXiv.org/abs/1404.3392}{\texttt{arXiv:1404.3392}}.
%%CITATION = ARXIV:1404.3392;%%.

\bibitem{Khachatryan:2014nda}
\hrefCMSnoop {}{{CMS Collaboration}, ``{Measurement of the ratio $\mathcal
  B({\rm t \to Wb})/\mathcal B({\rm t \to Wq})$ in pp collisions at $\sqrt{s}$
  = 8 TeV}'',} \textit{ Phys. Lett. B} \textbf{ 736} (2014) 33,
  \href{http://dx.doi.org/10.1016/j.physletb.2014.06.076}{\doi{10.1016/j.physletb.2014.06.076}},
\href{http://www.arXiv.org/abs/1404.2292}{\texttt{arXiv:1404.2292}}.
%%CITATION = ARXIV:1404.2292;%%.

\bibitem{Abazov:2006gd}
\hrefCMSnoop {}{{D0} Collaboration, ``{Evidence for production of single top
  quarks and first direct measurement of $|V_{\rm tb}|$}'',} \textit{ Phys.
  Rev. Lett.} \textbf{ 98} (2007) 181802,
  \href{http://dx.doi.org/10.1103/PhysRevLett.98.181802}{\doi{10.1103/PhysRevLett.98.181802}},
\href{http://www.arXiv.org/abs/hep-ex/0612052}{\texttt{arXiv:hep-ex/0612052}}.
%%CITATION = HEP-EX/0612052;%%.

\bibitem{Abazov:2008kt}
\hrefCMSnoop {}{{D0} Collaboration, ``{Evidence for production of single top
  quarks}'',} \textit{ Phys. Rev. D} \textbf{ 78} (2008) 012005,
  \href{http://dx.doi.org/10.1103/PhysRevD.78.012005}{\doi{10.1103/PhysRevD.78.012005}},
\href{http://www.arXiv.org/abs/0803.0739}{\texttt{arXiv:0803.0739}}.
%%CITATION = ARXIV:0803.0739;%%.

\bibitem{Aaltonen:2008sy}
\hrefCMSnoop {}{{CDF} Collaboration, ``{Measurement of the single-top-quark
  production cross section at CDF}'',} \textit{ Phys. Rev. Lett.} \textbf{ 101}
  (2008) 252001,
  \href{http://dx.doi.org/10.1103/PhysRevLett.101.252001}{\doi{10.1103/PhysRevLett.101.252001}},
\href{http://www.arXiv.org/abs/0809.2581}{\texttt{arXiv:0809.2581}}.
%%CITATION = ARXIV:0809.2581;%%.

\bibitem{Aaltonen:2009jj}
\hrefCMSnoop {}{{CDF} Collaboration, ``{Observation of electroweak single top
  quark production}'',} \textit{ Phys. Rev. Lett.} \textbf{ 103} (2009) 092002,
  \href{http://dx.doi.org/10.1103/PhysRevLett.103.092002}{\doi{10.1103/PhysRevLett.103.092002}},
\href{http://www.arXiv.org/abs/0903.0885}{\texttt{arXiv:0903.0885}}.
%%CITATION = 0903.0885;%%.

\bibitem{Aaltonen:2010jr}
\hrefCMSnoop {}{{CDF} Collaboration, ``{Observation of single top quark
  production and measurement of $|V_{\rm tb}|$ with CDF}'',} \textit{ Phys.
  Rev. D} \textbf{ 82} (2010) 112005,
  \href{http://dx.doi.org/10.1103/PhysRevD.82.112005}{\doi{10.1103/PhysRevD.82.112005}},
\href{http://www.arXiv.org/abs/1004.1181}{\texttt{arXiv:1004.1181}}.
%%CITATION = ARXIV:1004.1181;%%.

\bibitem{Abazov:2009ii}
\hrefCMSnoop {}{{D0} Collaboration, ``{Observation of single top-quark
  production}'',} \textit{ Phys. Rev. Lett.} \textbf{ 103} (2009) 092001,
  \href{http://dx.doi.org/10.1103/PhysRevLett.103.092001}{\doi{10.1103/PhysRevLett.103.092001}},
\href{http://www.arXiv.org/abs/0903.0850}{\texttt{arXiv:0903.0850}}.
%%CITATION = 0903.0850;%%.

\bibitem{Group:2009qk}
\hrefCMSnoop {}{{The CDF Collaboration, The D0 Collaboration, The Tevatron
  electroweak working group}, ``{Combination of CDF and D0 measurements of the
  single top production cross section}'',} (2009).
\href{http://www.arXiv.org/abs/0908.2171}{\texttt{arXiv:0908.2171}}.
%%CITATION = 0908.2171;%%.

\bibitem{Aaltonen:2014ura}
\hrefCMSnoop {}{{CDF} Collaboration, ``{Measurement of the single top quark
  production cross section and $|V_{\rm tb}|$ in events with one charged
  lepton, large missing transverse energy, and jets at CDF}'',} \textit{ Phys.
  Rev. Lett.} \textbf{ 113} (2014) 261804,
  \href{http://dx.doi.org/10.1103/PhysRevLett.113.261804}{\doi{10.1103/PhysRevLett.113.261804}},
\href{http://www.arXiv.org/abs/1407.4031}{\texttt{arXiv:1407.4031}}.
%%CITATION = ARXIV:1407.4031;%%.

\bibitem{Abazov:2013qka}
\hrefCMSnoop {}{{D0} Collaboration, ``{Evidence for $s$-channel single top
  quark production in p$\bar{\rm p}$ collisions at $\sqrt{s} = 1.96$ TeV}'',}
  \textit{ Phys. Lett. B} \textbf{ 726} (2013) 656,
  \href{http://dx.doi.org/10.1016/j.physletb.2013.09.048}{\doi{10.1016/j.physletb.2013.09.048}},
\href{http://www.arXiv.org/abs/1307.0731}{\texttt{arXiv:1307.0731}}.
%%CITATION = ARXIV:1307.0731;%%.

\bibitem{CDF:2014uma}
\hrefCMSnoop {}{{CDF, D0} Collaboration, ``{Observation of $s$-channel
  production of single top quarks at the Tevatron}'',} \textit{ Phys. Rev.
  Lett.} \textbf{ 112} (2014) 231803,
  \href{http://dx.doi.org/10.1103/PhysRevLett.112.231803}{\doi{10.1103/PhysRevLett.112.231803}},
  \href{http://www.arXiv.org/abs/1402.5126}{\texttt{arXiv:1402.5126}}.

\bibitem{Khachatryan:2016sib}
\hrefCMSnoop {}{{CMS Collaboration}, ``{Search for anomalous Wtb couplings and
  flavour-changing neutral currents in t-channel single top quark production in
  pp collisions at $\sqrt{s} =$ 7 and 8 TeV}'',} \textit{ JHEP} \textbf{ 02}
  (2017) 028,
  \href{http://dx.doi.org/10.1007/JHEP02(2017)028}{\doi{10.1007/JHEP02(2017)028}},
\href{http://www.arXiv.org/abs/1610.03545}{\texttt{arXiv:1610.03545}}.
%%CITATION = ARXIV:1610.03545;%%.

\bibitem{Aaboud:2019pkc}
\hrefCMSnoop {}{{ATLAS, CMS} Collaboration, ``{Combinations of single-top-quark
  production cross-section measurements and $|f_{\rm LV}V_{\rm tb}|$
  determinations at $\sqrt{s}=7$ and 8 TeV with the ATLAS and CMS
  experiments}'',} \textit{ JHEP} \textbf{ 05} (2019) 088,
  \href{http://dx.doi.org/10.1007/JHEP05(2019)088}{\doi{10.1007/JHEP05(2019)088}},
  \href{http://www.arXiv.org/abs/1902.07158}{\texttt{arXiv:1902.07158}}.

\bibitem{Lacker:2012ek}
H.~Lacker\hrefCMSnoop {}{ {et~al.}, ``Model-independent extraction of
  {$|V_{tq}|$} matrix elements from top-quark measurements at hadron
  colliders'',} \textit{ Eur. Phys. J. C} \textbf{ 72} (2012) 2048,
  \href{http://dx.doi.org/10.1140/epjc/s10052-012-2048-4}{\doi{10.1140/epjc/s10052-012-2048-4}},
\href{http://www.arXiv.org/abs/1202.4694}{\texttt{arXiv:1202.4694}}.
%%CITATION = ARXIV:1202.4694;%%.

\bibitem{AguilarSaavedra:2010wf}
\hrefCMSnoop {}{J.~A. Aguilar-Saavedra and A.~Onofre, ``{Using single top
  rapidity to measure $V_{\rm td}$, $V_{\rm ts}$, $V_{\rm tb}$ at hadron
  colliders}'',} \textit{ Phys. Rev. D} \textbf{ 83} (2011) 073003,
  \href{http://dx.doi.org/10.1103/PhysRevD.83.073003}{\doi{10.1103/PhysRevD.83.073003}},
\href{http://www.arXiv.org/abs/1002.4718}{\texttt{arXiv:1002.4718}}.
%%CITATION = ARXIV:1002.4718;%%.

\bibitem{Clerbaux:2018vup}
\hrefCMSnoop {}{B.~Clerbaux, W.~Fang, A.~Giammanco, and R.~Goldouzian,
  ``{Model-independent constraints on the CKM matrix elements $|V_{\rm tb}|$,
  $|V_{\rm ts}|$ and $|V_{\rm td}|$}'',} \textit{ JHEP} \textbf{ 03} (2019)
  022,
  \href{http://dx.doi.org/10.1007/JHEP03(2019)022}{\doi{10.1007/JHEP03(2019)022}},
\href{http://www.arXiv.org/abs/1807.07319}{\texttt{arXiv:1807.07319}}.
%%CITATION = ARXIV:1807.07319;%%.

\bibitem{Alvarez:2017ybk}
\hrefCMSnoop {}{E.~Alvarez, L.~Da~Rold, M.~Estevez, and J.~F. Kamenik,
  ``{Measuring $|V_{\rm td}|$ at the LHC}'',} \textit{ Phys. Rev. D} \textbf{
  97} (2018) 033002,
  \href{http://dx.doi.org/10.1103/PhysRevD.97.033002}{\doi{10.1103/PhysRevD.97.033002}},
\href{http://www.arXiv.org/abs/1709.07887}{\texttt{arXiv:1709.07887}}.
%%CITATION = ARXIV:1709.07887;%%.

\bibitem{Giammanco:2017xyn}
\hrefCMSnoop {}{A.~Giammanco and R.~Schwienhorst, ``{Single top-quark
  production at the Tevatron and the LHC}'',} \textit{ Rev. Mod. Phys.}
  \textbf{ 90} (2018) 035001,
  \href{http://dx.doi.org/10.1103/RevModPhys.90.035001}{\doi{10.1103/RevModPhys.90.035001}},
\href{http://www.arXiv.org/abs/1710.10699}{\texttt{arXiv:1710.10699}}.
%%CITATION = ARXIV:1710.10699;%%.

\bibitem{Chatrchyan:2008zzk}
\hrefCMSnoop {}{{CMS Collaboration}, ``The {CMS} experiment at the {CERN}
  {LHC}'',} \textit{ JINST} \textbf{ 3} (2008) S08004,
  \href{http://dx.doi.org/10.1088/1748-0221/3/08/S08004}{\doi{10.1088/1748-0221/3/08/S08004}}.

\bibitem{Khachatryan:2016bia}
\hrefCMSnoop {}{{CMS Collaboration}, ``{The CMS trigger system}'',} \textit{
  JINST} \textbf{ 12} (2017) P01020,
  \href{http://dx.doi.org/10.1088/1748-0221/12/01/P01020}{\doi{10.1088/1748-0221/12/01/P01020}},
\href{http://www.arXiv.org/abs/1609.02366}{\texttt{arXiv:1609.02366}}.
%%CITATION = ARXIV:1609.02366;%%.

\bibitem{Nason:2004rx}
\hrefCMSnoop {}{P.~Nason, ``A new method for combining {NLO QCD} with shower
  {M}onte {C}arlo algorithms'',} \textit{ JHEP} \textbf{ 11} (2004) 040,
  \href{http://dx.doi.org/10.1088/1126-6708/2004/11/040}{\doi{10.1088/1126-6708/2004/11/040}},
\href{http://www.arXiv.org/abs/hep-ph/0409146}{\texttt{arXiv:hep-ph/0409146}}.
%%CITATION = HEP-PH/0409146;%%.

\bibitem{Frixione:2007vw}
\hrefCMSnoop {}{S.~Frixione, P.~Nason, and C.~Oleari, ``{Matching NLO QCD
  computations with parton shower simulations: the POWHEG method}'',} \textit{
  JHEP} \textbf{ 11} (2007) 070,
  \href{http://dx.doi.org/10.1088/1126-6708/2007/11/070}{\doi{10.1088/1126-6708/2007/11/070}},
\href{http://www.arXiv.org/abs/0709.2092}{\texttt{arXiv:0709.2092}}.
%%CITATION = ARXIV:0709.2092;%%.

\bibitem{Alioli:2010xd}
\hrefCMSnoop {}{S.~Alioli, P.~Nason, C.~Oleari, and E.~Re, ``{A general
  framework for implementing NLO calculations in shower Monte Carlo programs:
  the POWHEG BOX}'',} \textit{ JHEP} \textbf{ 06} (2010) 043,
  \href{http://dx.doi.org/10.1007/JHEP06(2010)043}{\doi{10.1007/JHEP06(2010)043}},
\href{http://www.arXiv.org/abs/1002.2581}{\texttt{arXiv:1002.2581}}.
%%CITATION = 1002.2581;%%.

\bibitem{fourfiveflavorschemes}
\hrefCMSnoop {}{R.~Frederix, E.~Re, and P.~Torrielli, ``{Single-top $t$-channel
  hadroproduction in the four-flavour scheme with POWHEG and aMC@NLO}'',}
  \textit{ JHEP} \textbf{ 09} (2012) 130,
  \href{http://dx.doi.org/10.1007/JHEP09(2012)130}{\doi{10.1007/JHEP09(2012)130}},
  \href{http://www.arXiv.org/abs/1207.5391}{\texttt{arXiv:1207.5391}}.

\bibitem{Alioli:2009je}
\hrefCMSnoop {}{S.~Alioli, P.~Nason, C.~Oleari, and E.~Re, ``{NLO single-top
  production matched with shower in POWHEG: $s$- and $t$-channel
  contributions}'',} \textit{ JHEP} \textbf{ 09} (2009) 111,
  \href{http://dx.doi.org/10.1088/1126-6708/2009/09/111}{\doi{10.1088/1126-6708/2009/09/111}},
\href{http://www.arXiv.org/abs/0907.4076}{\texttt{arXiv:0907.4076}}.
%%CITATION = 0907.4076;%%.

\bibitem{Artoisenet:2012st}
\hrefCMSnoop {}{P.~Artoisenet, R.~Frederix, O.~Mattelaer, and R.~Rietkerk,
  ``Automatic spin-entangled decays of heavy resonances in {Monte Carlo}
  simulations'',} \textit{ JHEP} \textbf{ 03} (2013) 015,
  \href{http://dx.doi.org/10.1007/JHEP03(2013)015}{\doi{10.1007/JHEP03(2013)015}},
  \href{http://www.arXiv.org/abs/1212.3460}{\texttt{arXiv:1212.3460}}.

\bibitem{Frixione:2007nw}
\hrefCMSnoop {}{S.~Frixione, P.~Nason, and G.~Ridolfi, ``{A positive-weight
  next-to-leading-order Monte Carlo for heavy flavour hadroproduction}'',}
  \textit{ JHEP} \textbf{ 09} (2007) 126,
  \href{http://dx.doi.org/10.1088/1126-6708/2007/09/126}{\doi{10.1088/1126-6708/2007/09/126}},
\href{http://www.arXiv.org/abs/0707.3088}{\texttt{arXiv:0707.3088}}.
%%CITATION = ARXIV:0707.3088;%%.

\bibitem{Melia:2011tj}
\hrefCMSnoop {}{T.~Melia, P.~Nason, R.~R{\"o}ntsch, and G.~Zanderighi, ``{$\rm
  W^+W^-$, $\rm WZ$ and $\rm ZZ$} production in the {POWHEG BOX}'',} \textit{
  JHEP} \textbf{ 11} (2011) 078,
  \href{http://dx.doi.org/10.1007/JHEP11(2011)078}{\doi{10.1007/JHEP11(2011)078}},
\href{http://www.arXiv.org/abs/1107.5051}{\texttt{arXiv:1107.5051}}.
%%CITATION = ARXIV:1107.5051;%%.

\bibitem{Nason:2013ydw}
\hrefCMSnoop {}{P.~Nason and G.~Zanderighi, ``{$\rm W^+ W^-$, $\rm W Z$ and
  $\rm Z Z$} production in the {POWHEG-BOX-V2}'',} \textit{ Eur. Phys. J. C}
  \textbf{ 74} (2014) 2702,
  \href{http://dx.doi.org/10.1140/epjc/s10052-013-2702-5}{\doi{10.1140/epjc/s10052-013-2702-5}},
\href{http://www.arXiv.org/abs/1311.1365}{\texttt{arXiv:1311.1365}}.
%%CITATION = ARXIV:1311.1365;%%.

\bibitem{Re:2010bp}
\hrefCMSnoop {}{E.~Re, ``{Single-top $\rm Wt$-channel production matched with
  parton showers using the POWHEG method}'',} \textit{ Eur. Phys. J. C}
  \textbf{ 71} (2011) 1547,
  \href{http://dx.doi.org/10.1140/epjc/s10052-011-1547-z}{\doi{10.1140/epjc/s10052-011-1547-z}},
\href{http://www.arXiv.org/abs/1009.2450}{\texttt{arXiv:1009.2450}}.
%%CITATION = 1009.2450;%%.

\bibitem{amcatnlo}
J.~Alwall\hrefCMSnoop {}{ {et~al.}, ``The automated computation of tree-level
  and next-to-leading order differential cross sections, and their matching to
  parton shower simulations'',} \textit{ JHEP} \textbf{ 07} (2014) 079,
  \href{http://dx.doi.org/10.1007/JHEP07(2014)079}{\doi{10.1007/JHEP07(2014)079}},
  \href{http://www.arXiv.org/abs/1405.0301}{\texttt{arXiv:1405.0301}}.

\bibitem{Sjostrand:2007gs}
\hrefCMSnoop {}{T.~Sj{\"o}strand, S.~Mrenna, and P.~Z. Skands, ``{A brief
  introduction to PYTHIA 8.1}'',} \textit{ Comput. Phys. Commun.} \textbf{ 178}
  (2008) 852,
  \href{http://dx.doi.org/10.1016/j.cpc.2008.01.036}{\doi{10.1016/j.cpc.2008.01.036}},
\href{http://www.arXiv.org/abs/0710.3820}{\texttt{arXiv:0710.3820}}.
%%CITATION = ARXIV:0710.3820;%%.

\bibitem{Ctune}
\hrefCMSnoop {}{{CMS Collaboration}, ``{Event generator tunes obtained from
  underlying event and multiparton scattering measurements}'',} \textit{ Eur.
  Phys. J. C} \textbf{ 76} (2016) 155,
  \href{http://dx.doi.org/10.1140/epjc/s10052-016-3988-x}{\doi{10.1140/epjc/s10052-016-3988-x}},
\href{http://www.arXiv.org/abs/1512.00815}{\texttt{arXiv:1512.00815}}.
%%CITATION = ARXIV:1512.00815;%%.

\bibitem{CMS:2016kle}
\href {http://cdsweb.cern.ch/record/2235192}{{{CMS}} Collaboration,
  ``{Investigations of the impact of the parton shower tuning in PYTHIA 8 in
  the modelling of $\mathrm{t\overline{t}}$ at $\sqrt{s}=8$ and 13 TeV}'',} CMS
  Physics Analysis Summary CMS-PAS-TOP-16-021, 2016.

\bibitem{Frederix:2012ps}
\hrefCMSnoop {}{R.~Frederix and S.~Frixione, ``{Merging meets matching in
  MC@NLO}'',} \textit{ JHEP} \textbf{ 12} (2012) 061,
  \href{http://dx.doi.org/10.1007/JHEP12(2012)061}{\doi{10.1007/JHEP12(2012)061}},
\href{http://www.arXiv.org/abs/1209.6215}{\texttt{arXiv:1209.6215}}.
%%CITATION = ARXIV:1209.6215;%%.

\bibitem{NNPDF30}
\hrefCMSnoop {}{{NNPDF} Collaboration, ``Parton distributions for the {LHC Run
  II}'',} \textit{ JHEP} \textbf{ 04} (2015) 040,
  \href{http://dx.doi.org/10.1007/JHEP04(2015)040}{\doi{10.1007/JHEP04(2015)040}},
\href{http://www.arXiv.org/abs/1410.8849}{\texttt{arXiv:1410.8849}}.
%%CITATION = ARXIV:1410.8849;%%.

\bibitem{Agostinelli:2002hh}
\hrefCMSnoop {}{{GEANT4} Collaboration, ``{\GEANTfour}---a simulation
  toolkit'',} \textit{ Nucl. Instrum. Meth. A} \textbf{ 506} (2003) 250,
\href{http://dx.doi.org/10.1016/S0168-9002(03)01368-8}{\doi{10.1016/S0168-9002(03)01368-8}}.
%%CITATION = NUIMA,A506,250;%%.

\bibitem{fastjet2}
\hrefCMSnoop {}{M.~Cacciari, G.~P. Salam, and G.~Soyez, ``Fastjet user
  manual'',} \textit{ Eur. Phys. J. C} \textbf{ 72} (2012) 1896,
  \href{http://dx.doi.org/10.1140/epjc/s10052-012-1896-2}{\doi{10.1140/epjc/s10052-012-1896-2}},
\href{http://www.arXiv.org/abs/1111.6097}{\texttt{arXiv:1111.6097}}.
%%CITATION = ARXIV:1111.6097;%%.

\bibitem{Cacciari:2008gp}
\hrefCMSnoop {}{M.~Cacciari, G.~P. Salam, and G.~Soyez, ``The anti-\kt jet
  clustering algorithm'',} \textit{ JHEP} \textbf{ 04} (2008) 063,
  \href{http://dx.doi.org/10.1088/1126-6708/2008/04/063}{\doi{10.1088/1126-6708/2008/04/063}},
  \href{http://www.arXiv.org/abs/0802.1189}{\texttt{arXiv:0802.1189}}.

\bibitem{PFPAPER}
\hrefCMSnoop {}{{CMS Collaboration}, ``{Particle-flow reconstruction and global
  event description with the CMS detector}'',} \textit{ JINST} \textbf{ 12}
  (2017) P10003,
  \href{http://dx.doi.org/10.1088/1748-0221/12/10/P10003}{\doi{10.1088/1748-0221/12/10/P10003}},
\href{http://www.arXiv.org/abs/1706.04965}{\texttt{arXiv:1706.04965}}.
%%CITATION = ARXIV:1706.04965;%%.

\bibitem{Sirunyan:2018fpa}
\hrefCMSnoop {}{{CMS Collaboration}, ``{Performance of the CMS muon detector
  and muon reconstruction with proton-proton collisions at $\sqrt{s}=$ 13
  TeV}'',} \textit{ JINST} \textbf{ 13} (2018) P06015,
  \href{http://dx.doi.org/10.1088/1748-0221/13/06/P06015}{\doi{10.1088/1748-0221/13/06/P06015}},
\href{http://www.arXiv.org/abs/1804.04528}{\texttt{arXiv:1804.04528}}.
%%CITATION = ARXIV:1804.04528;%%.

\bibitem{Cacciari:2007fd}
\hrefCMSnoop {}{M.~Cacciari and G.~P. Salam, ``Pileup subtraction using jet
  areas'',} \textit{ Phys. Lett. B} \textbf{ 659} (2008) 119,
  \href{http://dx.doi.org/10.1016/j.physletb.2007.09.077}{\doi{10.1016/j.physletb.2007.09.077}},
  \href{http://www.arXiv.org/abs/0707.1378}{\texttt{arXiv:0707.1378}}.

\bibitem{Khachatryan:2016kdb}
\hrefCMSnoop {}{{CMS Collaboration}, ``{Jet energy scale and resolution in the
  CMS experiment in pp collisions at 8 TeV}'',} \textit{ JINST} \textbf{ 12}
  (2017) P02014,
  \href{http://dx.doi.org/10.1088/1748-0221/12/02/P02014}{\doi{10.1088/1748-0221/12/02/P02014}},
\href{http://www.arXiv.org/abs/1607.03663}{\texttt{arXiv:1607.03663}}.
%%CITATION = ARXIV:1607.03663;%%.

\bibitem{1748-0221-8-04-P04013}
\hrefCMSnoop {}{{CMS Collaboration}, ``{Identification of b-quark jets with the
  CMS experiment}'',} \textit{ JINST} \textbf{ 8} (2013) P04013,
  \href{http://dx.doi.org/10.1088/1748-0221/8/04/P04013}{\doi{10.1088/1748-0221/8/04/P04013}},
\href{http://www.arXiv.org/abs/1211.4462}{\texttt{arXiv:1211.4462}}.
%%CITATION = ARXIV:1211.4462;%%.

\bibitem{BTAGPAPER}
\hrefCMSnoop {}{{CMS Collaboration}, ``{Identification of heavy-flavour jets
  with the CMS detector in pp collisions at 13 TeV}'',} \textit{ JINST}
  \textbf{ 13} (2018) P05011,
  \href{http://dx.doi.org/10.1088/1748-0221/13/05/P05011}{\doi{10.1088/1748-0221/13/05/P05011}},
\href{http://www.arXiv.org/abs/1712.07158}{\texttt{arXiv:1712.07158}}.
%%CITATION = ARXIV:1712.07158;%%.

\bibitem{hator}
M.~Aliev\hrefCMSnoop {}{ {et~al.}, ``{HATHOR -- HAdronic Top and Heavy quarks
  crOss section calculatoR}'',} \textit{ Comput. Phys. Commun.} \textbf{ 182}
  (2011) 1034,
  \href{http://dx.doi.org/10.1016/j.cpc.2010.12.040}{\doi{10.1016/j.cpc.2010.12.040}},
\href{http://www.arXiv.org/abs/1007.1327}{\texttt{arXiv:1007.1327}}.
%%CITATION = ARXIV:1007.1327;%%.

\bibitem{Barlow:1993dm}
\hrefCMSnoop {}{R.~Barlow and C.~Beeston, ``{Fitting using finite Monte Carlo
  samples}'',} \textit{ Comput. Phys. Commun.} \textbf{ 77} (1993) 219,
\href{http://dx.doi.org/10.1016/0010-4655(93)90005-W}{\doi{10.1016/0010-4655(93)90005-W}}.
%%CITATION = CPHCB,77,219;%%.

\bibitem{bb_light}
\href {http://cdsweb.cern.ch/record/1306523}{J.~S. Conway, ``Nuisance
  parameters in likelihoods for multisource spectra'',} in \textit{
  {Proceedings of {PHYSTAT} 2011 Workshop on Statistical Issues Related to
  Discovery Claims in Search Experiments and Unfolding}}, H.~Prosper and
  L.~Lyons, eds., number CERN-2011-006, p.~115.
\newblock CERN, 2011.

\bibitem{Khachatryan:2010xn}
\hrefCMSnoop {}{{CMS Collaboration}, ``{Measurements of inclusive W and Z cross
  sections in pp collisions at $\sqrt{s}= $ 7 TeV}'',} \textit{ JHEP} \textbf{
  01} (2011) 080,
  \href{http://dx.doi.org/10.1007/JHEP01(2011)080}{\doi{10.1007/JHEP01(2011)080}},
\href{http://www.arXiv.org/abs/1012.2466}{\texttt{arXiv:1012.2466}}.
%%CITATION = ARXIV:1012.2466;%%.

\bibitem{Sirunyan:2018nqx}
\hrefCMSnoop {}{{CMS Collaboration}, ``{Measurement of the inelastic
  proton-proton cross section at $ \sqrt{s}=13 $ TeV}'',} \textit{ JHEP}
  \textbf{ 07} (2018) 161,
  \href{http://dx.doi.org/10.1007/JHEP07(2018)161}{\doi{10.1007/JHEP07(2018)161}},
\href{http://www.arXiv.org/abs/1802.02613}{\texttt{arXiv:1802.02613}}.
%%CITATION = ARXIV:1802.02613;%%.

\bibitem{Butterworth:2015oua}
\hrefCMSnoop {}{J.~Butterworth {et~al.}, ``{PDF4LHC recommendations for LHC run
  II}'',} \textit{ J. Phys. G} \textbf{ 43} (2016) 023001,
  \href{http://dx.doi.org/10.1088/0954-3899/43/2/023001}{\doi{10.1088/0954-3899/43/2/023001}},
\href{http://www.arXiv.org/abs/1510.03865}{\texttt{arXiv:1510.03865}}.
%%CITATION = ARXIV:1510.03865;%%.

\bibitem{CMS-PAS-LUM-17-001}
\href {http://cdsweb.cern.ch/record/2257069}{{{CMS}} Collaboration, ``{CMS}
  luminosity measurements for the 2016 data taking period'',} CMS Physics
  Analysis Summary CMS-PAS-LUM-17-001, 2017.

\bibitem{Abazov:2012vd}
\hrefCMSnoop {}{{D0} Collaboration, ``{Improved determination of the width of
  the top quark}'',} \textit{ Phys. Rev. D} \textbf{ 85} (2012) 091104,
  \href{http://dx.doi.org/10.1103/PhysRevD.85.091104}{\doi{10.1103/PhysRevD.85.091104}},
\href{http://www.arXiv.org/abs/1201.4156}{\texttt{arXiv:1201.4156}}.
%%CITATION = ARXIV:1201.4156;%%.

\bibitem{Aaboud:2017uqq}
\hrefCMSnoop {}{{ATLAS Collaboration}, ``{Direct top-quark decay width
  measurement in the $\rm t\bar{t}$ lepton+jets channel at $\sqrt{s}=8$ TeV
  with the ATLAS experiment}'',} \textit{ Eur. Phys. J. C} \textbf{ 78} (2018)
  129,
  \href{http://dx.doi.org/10.1140/epjc/s10052-018-5595-5}{\doi{10.1140/epjc/s10052-018-5595-5}},
\href{http://www.arXiv.org/abs/1709.04207}{\texttt{arXiv:1709.04207}}.
%%CITATION = ARXIV:1709.04207;%%.

\end{thebibliography}\endgroup
